%% file: main.tex
\documentclass[trackchanges,twocolumn,dvipsnames]{aastex702}
\usepackage{booktabs}
\usepackage{threeparttable}
\usepackage{multirow}
\usepackage{caption}
\usepackage{amsmath}

\newcommand{\Pan}{Pantheon+}
\newcommand{\Des}{DES-SN5YR}

\begin{document}

\newcommand{\jl}[1]{\textcolor{ForestGreen}{[{\bf JL}: #1]}}

\title{Supernovae Unite: Host-Galaxy Mass Measurements of Type Ia Supernovae and Their Impact on Cosmology}

\input{host_galaxy_author_list}







\begin{abstract}

Current consensus suggests that Type Ia supernova (SN Ia) brightnesses post light-curve standardization correlate with their host-galaxy stellar masses, which must be accounted for to obtain accurate cosmological constraints.~For example, \citet{vincenzi2025response} showed that different host-galaxy stellar mass measurements for the same dataset produce redshift-dependent differences of order $\sim 0.01$ mag, large enough to appreciably shift cosmological constraints.~We present internally consistent host-galaxy stellar masses remeasured using aperture photometry and spectral energy distribution (SED) fitting for SN-Unite, which combines the spectroscopic \Pan~and the photometric Dark Energy Survey five-year (\Des) samples into the largest SN Ia cosmology sample to date, with 2884 likely SNe Ia.~We find that photometry and SED fitting choices shift SN-Unite Flat$w$CDM parameters well below statistical uncertainties.~Our stellar masses differ from the \Pan~data release partly due to a redshift-dependent internal inconsistency within \Pan, while remaining largely consistent with the \Des~(DES-Dovekie) data release.~When the \Pan~subsample of SN-Unite is combined with Baryon Acoustic Oscillations (BAO) and Cosmic Microwave Background (CMB) measurements, the significance for time-evolving dark energy increases from 3.4$\sigma$ to 4.0$\sigma$ based on the maximum \textit{a posteriori} when our newly derived host-galaxy stellar masses replace the \Pan~data-release host-galaxy stellar masses, while DES-Dovekie remains virtually unchanged, consistent with the findings of \citet{hoyt2026union3.1_hosts}.~By remeasuring the host-galaxy stellar masses using a consistent framework throughout the whole sample, we improve the robustness of the SN-Unite cosmological constraints against systematic differences in host-galaxy stellar mass measurements. 






\end{abstract}

\keywords{\uat{Cosmology}{343} ---  \uat{Dark Energy}{351} --- \uat{Type Ia Supernovae}{1728} --- \uat{Galaxies}{573}  }


\section{Introduction}\label{sec:intro}

Type Ia Supernovae (SNe Ia) are powerful cosmological probes, especially in the context of dark energy. Not only did SNe Ia observations provide conclusive evidence of the accelerated expansion of the universe \citep{riess1998,perlmutter1999}, more than 25 years of observations \citep[e.g.,][]{Astier2006SNLSFirstYear,WoodVasey2007ESSENCE,Kessler2009SDSS2,sullivan2011snls3,Betoule2014JLA,brout2022pantheon+,DES-SN5YR,rubin2025union} have significantly tightened the constraints on the dark energy equation of state, $w = P/\rho$.~The preference for an accelerated expansion based on measurements of the deceleration parameter $q_0$ has accordingly increased from around $3\sigma$ in the late 1990s to over 5$\sigma$ with the recent Dark Energy Survey five-year analysis \citep[\Des,][]{DES-SN5YR}.  That trend continues with this current work, which is a companion paper to the SN-Unite cosmology paper that combines \Pan~and \Des~SNe Ia \citep{SN-Unite}. 

Furthermore, SNe Ia are independent probes of dark energy, providing complementary information to other probes of dark energy: the Cosmic Microwave Background (CMB) and Baryon Acoustic Oscillations (BAO). Therefore, combining all three measurements significantly tightens constraints on cosmological parameters.~While \Des~\citep{DES-SN5YR} reports $(\Omega_{\rm m}, w) = (0.264^{+0.074}_{-0.096},-0.80^{+0.14}_{-0.16})$ for SNe Ia alone constraints in Flat$w$CDM, the Dark Energy Spectroscopic Instrument Data Release 2 \citep[DESI DR2,][]{DESI_DR2_Cosmology} reports $(\Omega_{\rm m}, w) = (0.3098 \pm 0.0050,-0.971 \pm 0.021)$ when \Des~is combined with DESI BAO and CMB from \textit{Planck} \citep{Planck2018_I,Planck2018_VI}, the Atacama Cosmology Telescope (ACT) \citep{maccrann2024act,madhavacheril2024act,qu2024act}, and the South Pole Telescope (SPT) \citep{carlstrom2011SPT}. 

\begin{figure}
    \centering
    \includegraphics[width=0.95\linewidth]{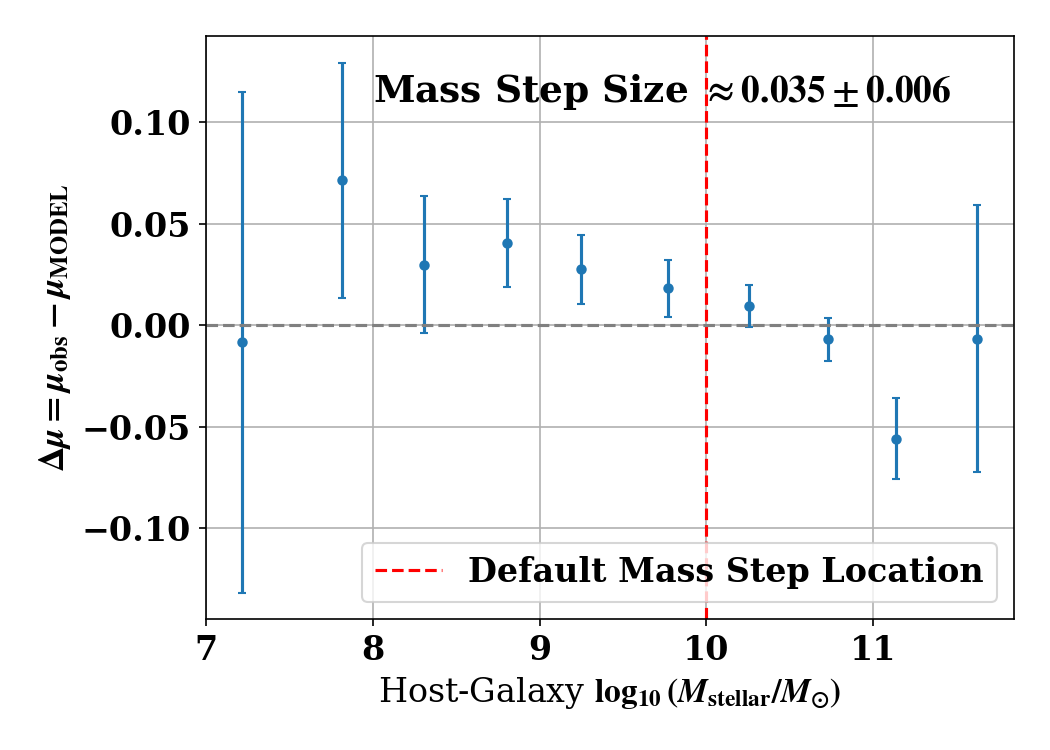}
    \caption{A visualization of host-galaxy stellar mass-step using the SN-Unite sample analyzed in this work.~There is a correlation between the Hubble residuals ($\Delta \mu$) and the host-galaxy stellar masses even after light-curve corrections. Here, $\Delta \mu$ is the Hubble residual with respect to the best-fit Flat$w$CDM cosmology. Host-galaxy stellar mass uncertainties ($\sim 0.2$ dex) are not shown for clarity.~For simplicity, $\Delta \mu$ error bars in this figure do not include the full covariance matrix.~Assuming a mass-step location of $10^{10} M_{\odot}$, we find the mass-step size to be approximately $0.035\pm 0.006$ from this figure.}
    \label{fig:mass-step_visualization}
\end{figure}

The use of SNe Ia as standardizable candles relies on the assumption that SNe Ia are consistent in their peak brightnesses, and that their intrinsic differences can be accounted for using empirical techniques.~One common method to standardize SNe Ia for cosmology is to fit for their light-curves using light-curve fitters such as the Spectral Adaptive Light Curve Template (SALT) \citep[SALT1 \& SALT2,][]{guy2005salt,guy2007salt2}, \citep[SALT3,][]{kenworthy2021salt3} to determine the light-curve parameters $x_1$ (`stretch') and $c$ (`color') for each SN Ia used for cosmological analysis.~After the stretch and color corrections, a residual correlation between SN Ia luminosities and host-galaxy properties remains. In Figure~\ref{fig:mass-step_visualization}, we show the correlation between the Hubble residuals and the host-galaxy stellar mass, when the host-galaxy stellar mass step is not accounted for (setting $\gamma =0$ in Equation~\ref{eq:tripp}) in the SN-Unite sample analyzed in this work.



The correlation was originally discovered as a dependence on host-galaxy stellar mass, with SNe Ia in higher stellar mass galaxies found to be 0.05 to 0.10 mag brighter than those in lower stellar mass galaxies after correcting for stretch and color \citep{kelly2010_mass-correlation,lampeitl2010hosts,sullivan2010_mass-step} ($> 4\sigma$ level for the latter two).~\citet{sullivan2010_mass-step} further found that including a host-galaxy stellar `mass step' improves the $\chi^2$ fit to the data, and that not accounting for the mass step can cause shifts in $w$ comparable to its statistical uncertainty in Flat$w$CDM.

Subsequent analyses found that alternative global host-galaxy properties show similar dependence - rest-frame color, emission lines, metallicity, age, star formation rate, as well as dust \citep[e.g.,][]{dandrea2011_hosts,gupta2011_hosts,moreno-raya2018,roman2018dependence,galbany2022_host,duarte2023sample,kelsey2023color,meldorf2023dust,wiseman2023age,martin2024_O_II,dixon2025_O_II}. Several works have additionally sought to characterize the correlation of SN Ia luminosities with local properties in the vicinity of the SNe Ia \citep{rigault2013environment,roman2018dependence,ginolin2025ztf_environment} due to concerns that global host-galaxy properties are not representative of the SN Ia environments. 

While some of the aforementioned alternative approaches are more physically motivated, there is not yet a clear consensus that such approaches provide a more robust correction than the global host-galaxy stellar mass.~Moreover, global host-galaxy stellar masses are generally more straightforward to measure accurately and precisely across large, heterogeneous SN Ia samples.~As such, host-galaxy stellar masses have continued to be used for SN Ia cosmological analyses to account for host-dependence of SNe Ia luminosities, including in three of the most widely utilized recent SN Ia analyses: \Pan~\citep{brout2022pantheon+}, \Des~\citep{DES-SN5YR}, and Union3 \citep{rubin2025union}.~Therefore, we limit our scope to host-galaxy stellar masses rather than other host-related parameters in this work.

~In \Des, the full Tripp equation \citep{Tripp1998} for each SN Ia $i$ is given as: 
\begin{equation}
    \mu_{{\rm obs},i} =m_{x,i} +\alpha x_{1,i} -\beta c_{i}+\gamma G_{{\rm host},i} -M - \Delta \mu_{\rm bias}, \label{eq:tripp}
\end{equation}

\noindent with $\gamma G_{\rm host}$ used to capture the dependency between the SN Ia luminosity and host-galaxy property (mass in the nominal analysis), with $G_{\rm host} = \pm \frac{1}{2}$ depending on whether the host is above or below the host-galaxy stellar mass step.~\Des~found $\gamma = 0.038 \pm 0.007$, or more than 5$\sigma$ significance that there \textit{is} a host-galaxy mass step at $10^{10} M_{\odot}$.~The DES-Dovekie reanalysis of \Des~\citep{dovekie2025} found a slightly smaller significance at $\gamma = 0.033 \pm 0.008$. 

While recent surveys such as \Des~and \Pan, as well as earlier surveys like the Supernova Legacy Survey (SNLS) and the Joint Light-curve Analysis (JLA)~\citep{Conley2010SNLS3,sullivan2011snls3,Betoule2014JLA}, used a `mass step' to account for the host-galaxy mass, other analyses have considered smooth functions and the redshift evolution of the `mass step' \citep{vincenzi2024DES_systematics,dovekie2025,rubin2025union,rubin2026union3.1}. 

\citet{smith2020hosts} additionally explores the systematic uncertainties of their host-galaxy stellar mass estimates for the 206 spectroscopically confirmed Dark Energy Survey `three-year' \citep[DES-SN3YR,][]{DES-SN3YR} analysis.~They find that their host mass estimates are robust to systematic variants such as introducing additional bursts in star-formation history or using a different initial mass function (IMF), meaning that such variants can result in an overall systematic shift by 0.1 to 0.5 dex, but not a stellar mass-dependent offset.


Since the report of the preference of time-evolving dark energy over Flat$\Lambda$CDM \citep{DES-SN5YR,DESI_DR1_Cosmology,DESI_DR2_Cosmology} with $>3\sigma$ significance when the DESI DR1/DR2 BAO are combined with the CMB and the \Des~or Union3 datasets, SN Ia cosmology has been under increased scrutiny \citep{efstathiou2025_evolving_or_systematics,notari2025}.~This is partly because the best-fit Flat$w_0 w_a $CDM results under the linear Chevallier-Polarski-Linder~(CPL) parametrization \citep[where $w=w_0 + w_a(1-a)$,][]{chevallier01, Linder_2003} are especially sensitive to low-$z$ ($z < 0.2$) SN Ia data where other cosmological constraints are largely absent. 

\citet{efstathiou2025_evolving_or_systematics} in particular, raised the possibility that the increased preference for time-evolving dark energy when \Des~($3.9\sigma$ in DESI DR1 based on the maximum \textit{a posteriori} (MAP)) is used instead of \Pan~($2.5\sigma$ in DESI DR1 based on the MAP) to combine with DESI DR1 BAO and the CMB, was due to unaddressed systematics in the \Des~analysis, leading to a $\sim 0.04$ mag difference between the low-$z$ and high-$z$ overlapping samples of \Des~and \Pan.~In a response to \citet{efstathiou2025_evolving_or_systematics}, \citet{vincenzi2025response} found that these differences are justified in part because the photometrically classified \Des~sample is substantially more complete and subject to weaker selection effects than the spectroscopically selected DES subset included in \Pan. Additional differences are due to improved intrinsic scatter modeling and improved host-galaxy stellar mass measurements.~The difference in host-galaxy mass measurements contributes to $\sim 0.01$ mag out of the $\sim 0.04$ mag difference confirmed in \citet{vincenzi2025response}. 

More recently, the DES-Dovekie analysis \citep{dovekie2025}, a reanalysis of \Des~with an updated SN Ia photometric cross-calibration, found a significance for time-evolving dark energy at $3.2\sigma$ when combined with DESI-DR2 BAO \citep{DESI_DR2_Cosmology} and the CMB based on the maximum \textit{likelihood} (ML), down from $4.2\sigma$ (based on MAP) when \Des~is combined instead.~Intriguingly, the Union3.1 analysis \citep{rubin2026union3.1,hoyt2026union3.1_hosts} finds $3.4\sigma$ (based on MAP), when Union3.1 (using UNITY1.7 or UNITY1.8) SNe Ia are combined with DESI-DR2 and \textit{Planck} CMB, down from $3.8\sigma$ with Union3 \citep{rubin2025union}.~Furthermore, \citet{hoyt2026union3.1_hosts} find that updating \Pan~with newly re-derived host-galaxy stellar masses to correct for a redshift-dependent internal inconsistency in the original \Pan~analysis increases the preference for time-evolving dark energy from $2.8\sigma$ (reported in DESI-DR2) to $3.2\sigma$, again, when combined with DESI-DR2 BAO and \textit{Planck} CMB. 

While the \citet{hoyt2026union3.1_hosts} reanalysis of \Pan~with updated host-galaxy stellar masses was not done with the same analysis framework as \Pan~\citep{brout2022pantheon+,scolnic2022pantheon+}, the concordance between the three currently most often utilized SN Ia datasets, in terms of the preference for time-evolving dark energy when DESI-DR2 BAO and \textit{Planck} CMB are combined, is striking. 

For some context, \Pan~and Union3/Union3.1 share a similar dataset, using a compilation of historical samples of spectroscopically confirmed SNe Ia, while \Des~and DES-Dovekie mostly use photometrically classified SNe Ia from the DES-SN program \citep{moller2020snn,qu2021scone,moller2022,vincenzi2023des_classification_biases,sanchez2024DES_DR}, with a small ($\sim 200$) external low-$z$ sample that is largely a subset of the \Pan~and Union low-$z$ datasets.~All three datasets have a similar number of SNe Ia (around 1500 to 2000) and therefore similar statistical constraining powers on cosmological parameters.~For the methodology, \Pan~and \Des~(DES-Dovekie) use the same framework, which uses Beams with Bias Corrections \citep[BBC,][]{2017_BiasCor} to perform simulations-based corrections to model selection effects, while Union uses the Unified Nonlinear Inference for Type Ia cosmologY  \citep[UNITY,][]{rubin2015unity,rubin2025union}, a Bayesian hierarchical model that simultaneously models the SN Ia light-curve parameters, selection effects and the cosmological parameters. 

In the light of such tantalizing developments, SN-Unite \citep{SN-Unite} combines the \Des~and \Pan~datasets to create the largest SN Ia dataset to date numbering $2884$ likely SNe Ia ($P_{\rm Ia} > 0.8$, with about 46\% of SNe Ia spectroscopically confirmed), and provides the tightest constraints to date on $\Omega_m$ and dark energy using SNe Ia only, as well as with combinations of SNe Ia, BAO, and the CMB. 

Due to the developments elucidated above, it is crucial that we investigate host-galaxy mass measurements and their impact on cosmology in detail.~For this reason, we remeasure the host-galaxy stellar masses for the SN-Unite sample with a single, consistent pipeline, and assess how measurement choices impact the SN-Unite as well as the cosmological analyses of the \Pan~and \Des~subsamples within SN-Unite.

This work is organized as follows.~In Section~\ref{sec:data}, we describe our dataset and in Section~\ref{sec:methods}, we describe our methodology, including our photometry pipeline, as well as the spectral energy distribution (SED) fitting we use to estimate the host-galaxy stellar mass.~In Section~\ref{sec:results_mass}, we show our host-galaxy stellar mass measurements, and compare with previous measurements in the literature.~We describe and assess the impact of different methodology choices on cosmology in Section~\ref{sec:results_cosmo_ana_choice}.~In Section~\ref{sec:results_cosmo_DR_masses}, we show cosmological constraints for the \Des~and \Pan~subsamples within SN-Unite and assess the impact of updating the data-release masses with SN-Unite masses, in Flat$w$CDM using SNe Ia only, and in Flat$w_0w_a$CDM using SNe Ia + BAO + CMB. We end with a Discussion and Conclusion in Section~\ref{sec:concl}.

\section{Data} \label{sec:data}

In this section, we describe the datasets used for the SN-Unite host-galaxy stellar mass measurements. 

\subsection{The SN-Unite sample} \label{sec:data_Unite}

We re-determine the host-galaxy masses for over 98\% of the SNe Ia in the SN-Unite parent sample, that is, 5271 SNe Ia that have host-galaxy spectroscopic redshifts and converged SALT3 light-curve fits (for the list, see Table 3 of \citealt{SN-Unite}).~For $< 2\%$ of the parent sample, we are not able to re-determine the photometry, and hence host stellar masses.~This is typically due to suboptimal image quality or the hosts being too faint for accurate photometry with our pipeline outlined in Section~\ref{sec:methods_photometry}.~We release the host-galaxy properties including the spectroscopic redshifts and Right Ascension (RA) and Declination (Dec)\footnote{For 143 or $< 3\%$ of the \Pan~host-galaxies, the host-galaxy coordinates are not available, so we use the SN Ia RA and Dec instead.} adopted from the \Pan\footnote{\url{https://github.com/PantheonPlusSH0ES/DataRelease}} and \Des/DES-Dovekie\footnote{\url{https://github.com/des-science/DES-SN5YR}} data-release, as well as our remeasured host-galaxy stellar masses and photometry with our main data-release at: \url{https://github.com/jasonlee17/SN-Unite_Host_Galaxy_Masses}.~Note that DES-Dovekie adopted all host-galaxy photometry and properties from \Des, so we use the two interchangeably in subsequent sections unless we are discussing cosmological constraints.~All of the host-galaxies in the SN-Unite sample use some form of the directional-light-radius (DLR) method to match SNe Ia with their host-galaxies \citep{sullivan2006_hosts,gupta2016hosts}, and \citet{qu2024host-mismatch} demonstrate that the mis-identification of hosts has a negligible impact on \Des~cosmology, with the mis-identification rate being less than 2\%.~Additionally, we inspect each individual galaxy image as highlighted in Section~\ref{sec:methods_photometry}, which reduces the risk of host-misidentification even further.

For the majority of this analysis, we focus on the measurements on the \textbf{Cosmology sample}, which are the SNe that are included in the SN-Unite Hubble Diagram (see Fig.~1 of \citet{SN-Unite}) used for cosmological inference.~The Cosmology sample consists of 2884 SN Ia observations after selection cuts, corresponding to 2742 unique SNe Ia, with some SNe Ia observed by multiple surveys.~The number of unique host-galaxies is slightly smaller, as multiple SNe Ia can occasionally occur in the same galaxy.~For the 40 ($< 2\%$) host-galaxies in the Cosmology sample where we are not able to obtain photometry, we adopt the \Pan~or \Des~data-release values, which has a negligible impact in our analysis.~The sample consists of host-galaxies of SN Ia candidates for:

\begin{itemize}
    \item 1549 discovered by DES-SN3YR and \Des~\citep{Brout_2019, Smith_2020survey,sanchez2024DES_DR,vincenzi2024DES_systematics,DES-SN5YR}
    \item 251 by the Pan-STARRS1 Medium-Deep Survey \citep[PS1MD,][]{Scolnic2018Pantheon}
    \item 171 by the Supernova Legacy Survey \citep[SNLS,][]{Astier2006SNLSFirstYear,Conley2010SNLS3,sullivan2011snls3,Betoule2014JLA}
    \item 10 by the \textit{Hubble Space Telescope} \citep[HST,][]{Gilliland_1999, Riess_2001, Riess_2004, Riess_2007, Suzuki2012Union2.1, Riess_2018} 
    \item 256 by the Sloan Digital Sky Survey \citep[SDSS,][]{sako2018SDSS}
    \item and 647 by various low-$z$ surveys, including the Lick Observatory Supernova Search \citep[LOSS1, LOSS2,][]{Ganeshalingam_2010,stahl_2019}, the Swift Optical/Ultraviolet Supernova Archive \citep[SWIFT,][]{Brown2014}, A Complete Nearby (Redshift $<$ 0.02) Sample of Type Ia Supernova Light Curves \citep[CNIa0.02,][]{Chen_2022}, the Carnegie Supernova  Project \citep[CSP,][]{Krisciunas_2017}, and four Center for Astrophysics SN surveys \citep[CfA1, CfA2, CfA3, CfA4,][]{Riess_1999,Jha_2006,Hicken_2009_CfA3,Hicken_2012_CfA4}, and various others \citep{Zhang_2009, Milne_2010, Stritzinger_2010, tsvetkov2010sn2008fvtypeia, Gall_2018, Burns_2018, Burns_2020, Kawabata_2020}
\end{itemize}

\subsection{Photometry Dataset} \label{sec:data_photometry}

Although high-quality host-galaxy spectra can provide more precise and accurate host-galaxy properties, spectra are not uniformly available for the SN-Unite host-galaxies.~In addition, the spectra are often not of high enough quality for spectral fitting since the primary purpose of taking spectra (at least for the case of DES) was to obtain host redshifts.~Hence the first step in measuring host-galaxy properties is to obtain photometry of the host-galaxies. In this work, we use optical and near-ultraviolet observations for our default host-galaxy stellar masses. For the optical wavelengths, we use the DES \textit{griz} bands from 5-year median coadded images as in \citet{qu2024host-mismatch} for the \Des~hosts.~While \citet{wiseman2020_DES_COADD} (W20) compiled DES \textit{griz} host-galaxy photometry using 4-year (excluding the season of SN Ia explosion) coadded images for DES-SN5YR, we re-derive the photometry for SN-Unite to maintain a consistent photometry pipeline throughout all of the host-galaxy sample.~We note that median coadded images that include epochs containing SN Ia light can retain a small positive flux bias at pixels within the SN Ia PSF, as the SN Ia flux shifts the distribution of the contributing pixel values toward higher values and can therefore increase the sample median.~In Table~\ref{tab:w20_coadd_unite_photometry}, we compare the W20 photometry with SN-Unite photometry for \Des~hosts and find that the difference is quite small, around $0.015$ mag or less, well below the threshold needed to impact the host-galaxy stellar mass measurements presented in this work.~For the \Pan~sample, we use coadded images from: 

\begin{table}
\centering
\caption{
Comparison of host-galaxy photometry between \citet{wiseman2020_DES_COADD} COADD measurements and the median coadded images of \textsc{SN-Unite}.~While median coadded images can be biased, this table illustrates that the bias is less than $0.015 $ mag across all bands, which does not impact our host-galaxy stellar mass measurements.~Differences are defined as
$\Delta m = m_{\mathrm{W20,COADD}} - m_{\mathrm{SN-UNITE}}$, while quoted values are the mean, error on the mean, and rms of $\Delta m$. 
Quoted values are computed after removing catastrophic outliers using a
$5\sigma$ clip based on the median absolute deviation of $\Delta m$. The rms is computed from the clipped $\Delta m$ distribution in each band.
}
\label{tab:w20_coadd_unite_photometry}
\begin{tabular}{lcc}
\hline
Band &
$\langle \Delta m \rangle$ (rms) &
$N_{\mathrm{used}}/N_{\mathrm{tot}}$ \\
\hline

\textit{g} & $-0.001 \pm 0.003$ (0.119) & 1428/1509 \\
\textit{r} & $0.014 \pm 0.002$ (0.074) & 1409/1512 \\
\textit{i} & $0.010 \pm 0.002$ (0.073) & 1400/1512 \\
\textit{z} & $0.010 \pm 0.002$ (0.070) & 1399/1512 \\

\hline
\end{tabular}
\\[2mm]
\end{table}

\begin{enumerate}
    \item the broader DES Data Release 2 \citep[DES-DR2,][]{abbott2021DES_DR2} \textit{griz} bands ($\sim$5000 $\rm{deg}^2$)
    \item the LegacySurvey Data Release 10 \citep[LegacySurvey-DR10,][]{DESI_LegacySurvey2019}, which consists of the Mayall $z$-band Legacy Survey (MzLS), Dark Energy Camera Legacy Survey (DECaLS), and the Beijing-Arizona Sky Survey (BASS) - \textit{griz} for DECaLS, and \textit{grz} for BASS ($\sim$ 20000 $\rm{deg}^2$) for \textit{grz} and $\sim$ 18000 $\rm{deg}^2$ for \textit{i})
    \item the PanSTARRS \textit{griz} bands \citep{magnier2020PS} ($\sim$ 30000 $\rm{deg}^2$ at $\rm{Dec} > -30^{\circ}$)
    \item the SkyMapper Southern Survey \citep[SMSS,][]{onken2024skymapper} \textit{uvgriz} bands ($\sim$ 26000 $\rm{deg}^2$ at $\rm{Dec} < 16^{\circ}$) only for the 5 $H_0$ calibrators (2 unique hosts) outside the footprints of DES-DR2, LegacySurvey-DR10, and PanSTARRs,
\end{enumerate}
\noindent with images from survey 2 being used when the host is outside the footprint of survey 1 and images from survey 3 being used when the host is outside of the footprint of survey 2.

Out of the 2844 hosts for which masses were determined in this work, optical photometry for 1519 hosts was obtained using 5-year median DES coadds (\Des), 310 using DES-DR2 coadds, 823 using LegacySurvey images, 187 using PanSTARRS images, and 5 using SMSS.

A major improvement from the \Pan~data-release host-galaxy properties is the use of significantly deeper DES and LegacySurvey-DR10 images as opposed to PanSTARRS or SDSS images for the majority of the \Pan~host-galaxies. We only use PanSTARRS images when the hosts are not in the DES or LegacySurvey-DR10 footprints, or if there are image quality issues with LegacySurvey images (e.g., bad pixels). The 5$\sigma$ depths of each set of images are shown in Table~\ref{tab:survey_depths}. 

\begin{table*}
\centering
\begin{tabular}{lccccc}
\hline
Survey & $g$ & $r$ & $i$ & $z$ & Notes \\
\hline

DES-SN5YR median coadds (shallow) & 26.3 & 26.0 & 25.7 & 25.5 & Computed in this work using 1.95$\arcsec$ apertures \\ 

DES-SN5YR median coadds (deep) & 27.2 & 27.2 & 26.9 & 26.7 & Computed in this work using 1.95$\arcsec$ apertures\\

DES-DR2 & 25.5 & 25.2 & 24.6 & 23.9 & Converted from published $10\sigma$ 1.95$\arcsec$ aperture depths \\

LegacySurvey-DR10 & 24.7 & 23.6 & --- & 22.8 & Typical $5\sigma$ point-source coadd depths\\

PanSTARRS (3$\pi$) & 23.3 & 23.2 & 23.1 & 22.3& Published mean $5\sigma$ stacked point-source depths \\
\hline
\end{tabular}
\caption{Typical 5$\sigma$ point-source depths (AB magnitudes) for wide-field imaging surveys used in host-galaxy photometry. Depths correspond to stacked survey coadds and vary across the footprint.~Note that DES-SN5YR median coadds are separated into shallow and deep fields, with roughly 70\% of host-galaxies in the DES-SN5YR subsample being from the shallow fields, and 30\% being from the deep fields.~Note that the LegacySurvey-DR10 does not report \textit{i}-band depths.}
\label{tab:survey_depths}
\end{table*}

We additionally considered the risk of SNe Ia light biasing host-property estimates.~The \textit{median} coadded images from \Des~exclude SNe Ia light by definition, and optical survey images used for the \Pan~sample are also from coadds of at least 4-years.~For the $\sim200$ host galaxies that are found in DES, LegacySurvey-DR10, and PanSTARRS, the optical photometry and hence the derived host-galaxy stellar masses were found to be consistent.~This further suggests that risk for SNe Ia contamination is minimal, as the three surveys were observed during different years. 

We also use the \textit{NUV} images from the \textit{Galaxy Evolution Explorer} \citep[\textit{GALEX},][]{martin2005GALEX} when available, which is over $90\%$ of our host-galaxies.~While \textit{JHKs} images from the \textit{2 Micron All Sky Survey} \citep[\textit{2MASS},][]{skrutskie2006_2MASS} are also widely used, we find that adding \textit{2MASS} \textit{JHKs} photometry degrades the quality of the spectral energy distribution (SED) fit described in Section~\ref{sec:SED_fitting} for the majority of our host-galaxies, likely due to the near-IR image resolution being suboptimal for our analysis, with the \textit{JHKs} photometry scatter being around 0.8 mag.~Nonetheless, we include the \textit{2MASS} \textit{JHKs} photometry when considering variants of our mass measurements in Sections~\ref{sec:results_mass} and~\ref{sec:results_cosmo}. We find that the addition of the near-UV photometry reduces the median host-galaxy stellar mass uncertainty by about 10\% (from $0.25$ dex to $0.23$ dex, see left plot of Figure~\ref{fig:log_mass_err_distributions} in Appendix~\ref{sec:appendixA}), but does not change the median host-galaxy log stellar mass measurements, as shown in Table~\ref{tab:unite_mass_variants}.  

We note that our wavelength coverage of $180 - 1000$ nm at $z = 0$ does not probe the same parts of the galaxy spectra at higher redshifts.~At $z > 0.5$, we are no longer sensitive to any part of the rest-frame near-IR, and at $z > 1$, we are only sensitive to rest-frame UV and blue/green optical wavelengths.~\citet{paulino2022_optical_SED} find that inconsistent rest-frame coverage in photometry can cause up to a $\sim0.3$ dex underestimation of host-galaxy stellar masses at $z \sim 1$, but shifts cosmological constraints by less than 1\%.~Additionally, we find that not including the near-IR (or near-UV) has a small impact on our mass measurements (see Table~\ref{tab:unite_mass_variants} and Figure~\ref{fig:Unite_vs_Mass_variants} for more details), although other host properties are more sensitive to the wavelength coverage.


\section{Methodology} \label{sec:methods}

We describe our host-galaxy stellar mass estimation pipeline in this section. 

\subsection{Photometry} \label{sec:methods_photometry}

As mentioned in Section~\ref{sec:data_photometry}, we start with photometry of the host-galaxies.~We use the \verb|HostPhot| (v3.1.3)\footnote{https://github.com/temuller/hostphot} \citep{muller2022hostphot} package to perform aperture photometry using host RA and Dec coordinates obtained by \Des, SNLS, PS1MD, SDSS, and \Pan, largely based on the directional-light-radius (DLR) approach to identify host-galaxies \citep{sullivan2006_hosts,gupta2016hosts}.~While model-fitting (e.g., fitting a Sersic profile) as in \citet{qu2024host-mismatch} is also possible, we perform aperture photometry as done in \Pan~and DES \citep{wiseman2020_DES_COADD} in order to keep the photometry pipeline consistent throughout all redshifts.

~Host-galaxy photometry with \verb|HostPhot| begins with downloading image cutouts; we use $3 \arcmin$ by $3 \arcmin$ cutouts for $z_{\rm helio} < 0.03$ hosts (larger for the nearest galaxies), $1.5 \arcmin$ by $1.5 \arcmin$ cutouts for $0.03 \le z_{\rm helio} < 0.10$ hosts, and $0.5 \arcmin$ by $0.5 \arcmin$ cutouts for $ z_{\rm helio} \ge 0.10$ hosts where $z_{\rm helio}$ denotes the heliocentric redshift.~For the DES 5-year median coadds, we use a custom code to obtain 100 pixel by 100 pixel (roughly $0.45 \arcmin$ by $0.45 \arcmin$) cutouts.~For reference, $3 \arcmin$ corresponds to roughly 108 kpc at $z = 0.03$, while $1.5 \arcmin$ corresponds to roughly 166 kpc at $z = 0.1$, and $0.5 \arcmin$ corresponds to roughly 55 kpc at $z = 0.1$ assuming a Flat$\Lambda$CDM cosmology with $H_0 = 70 $km/s/Mpc and $\Omega_m = 0.3$.~We use \textit{riz} (\textit{grz} when \textit{i} is not available) coadded images to determine the aperture used for photometry in all bands. We set  \verb|optimize_kronrad| to \verb|True| and set the default \verb|eps| to 0.001, which means that the Kron radius is increased until the change in flux is less than 0.1\%.~We also use the masking functionality to mask contaminants.~We keep the aperture and masking settings the same for optical, near-UV, and near-IR.

\begin{table}[]
    \centering
    \begin{tabular}{cc}
    \toprule
    Success Scores &  Meaning \\
    \midrule
    1.0     &     Successful \\
    0.7    &   Minor contamination \\
    0.5   &  Some issues but not enough to change \\ & mass step bin   \\
    0.3   &  Major issues (rerun) \\
    0   &  Incorrect host (rerun) \\
    \bottomrule
    \end{tabular}
    \caption{Success scores assigned to coadded images for aperture determination.~When the success score is 0.3 or 0, we tweak HostPhot with different choices (e.g., smaller, larger aperture, or no masking) then rerun.~We only use host-galaxy mass measurements with a success score of 0.5 or higher. }
    \label{tab:success_scores}
\end{table}

\begin{figure*}
    \centering
    \includegraphics[width=0.33\linewidth]{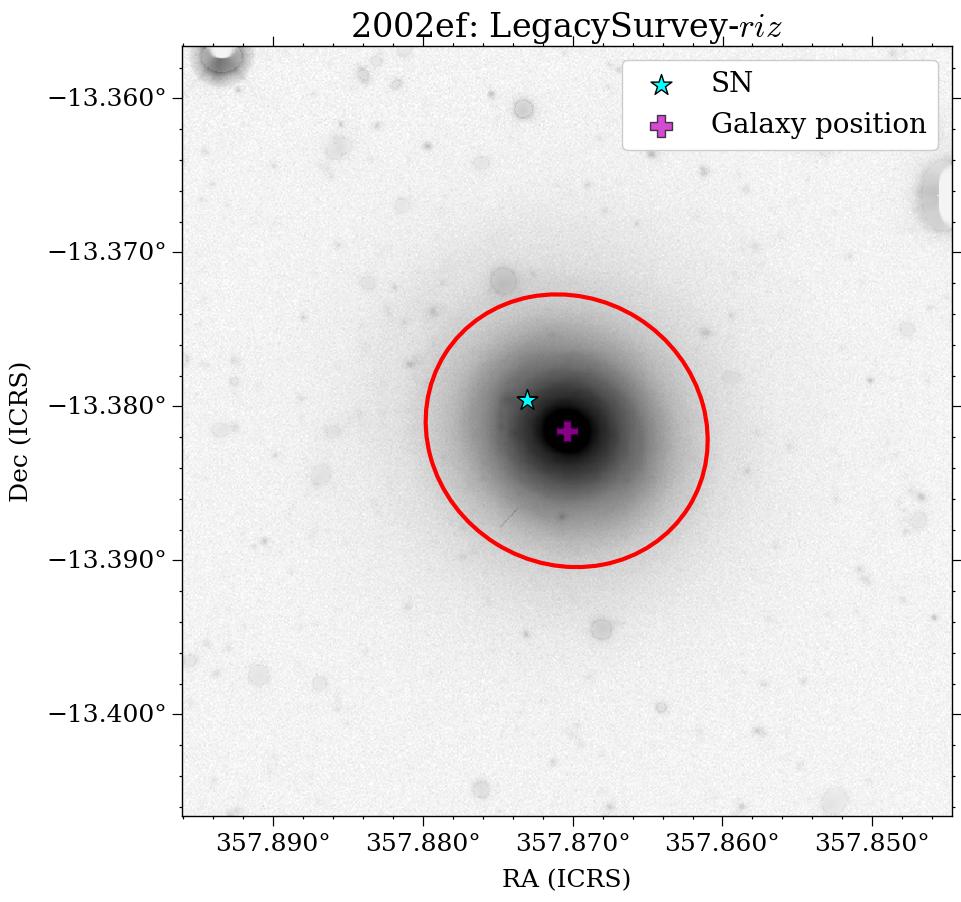}\includegraphics[width=0.33\linewidth]{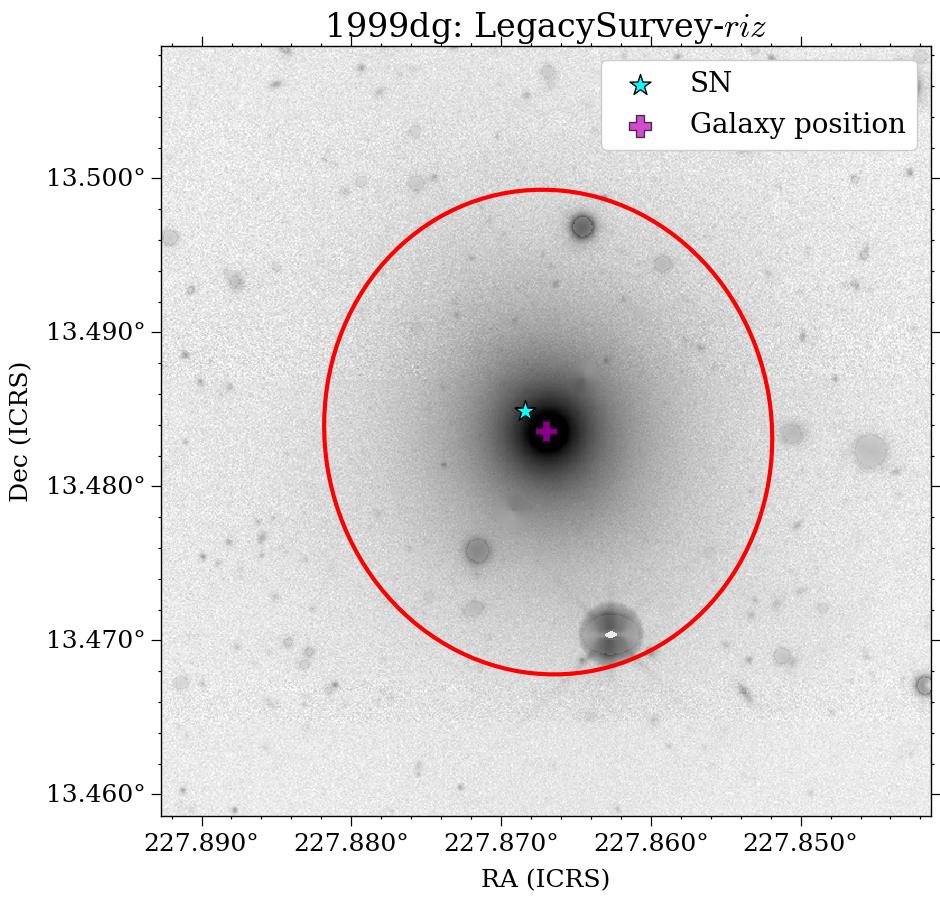}\includegraphics[width=0.33\linewidth]{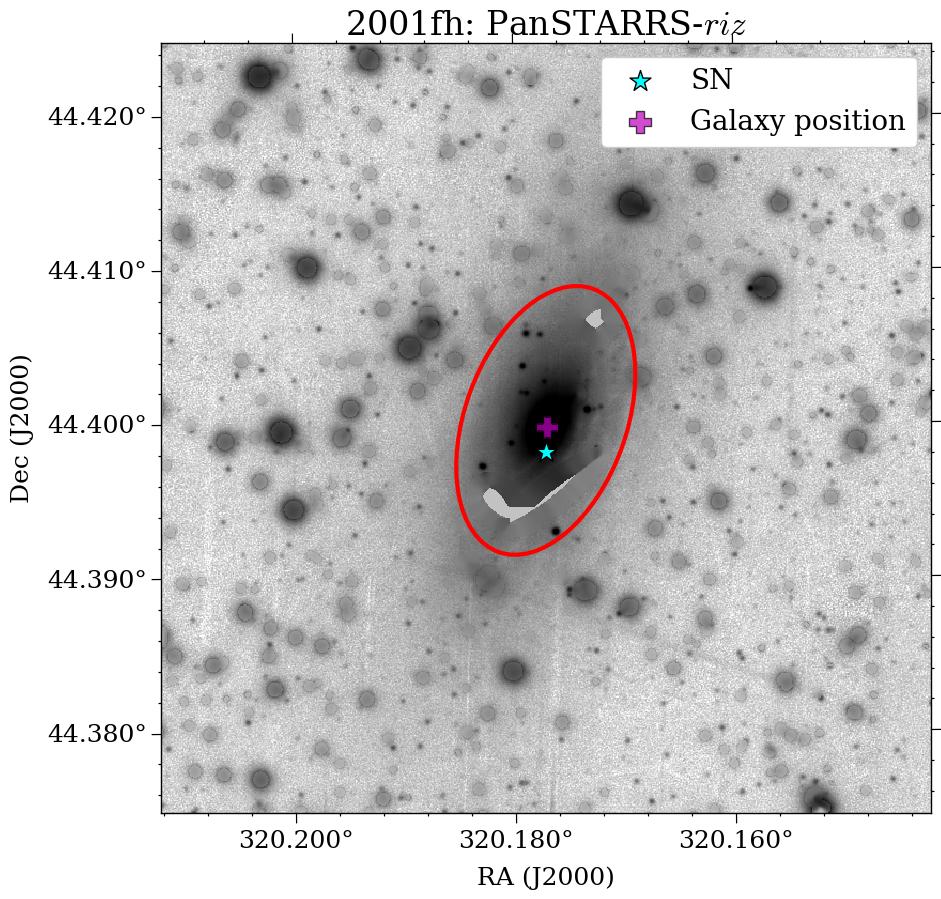}
    \caption{Examples of success scores given to HostPhot apertures for three very low-$z$ ($z < 0.03$) SN Ia host-galaxies: (Left) 1.0, or successful (Middle) 0.7, with minor contamination (Right) 0.5, with considerable contamination, but not enough to change mass step bins.~From left to right, $z_{\rm HD} = 0.0227$, $\log_{10}{(M_{\rm stellar}/M_{\odot})} = 10.79^{+0.15}_{-0.15}$; $z_{\rm HD} = 0.0220$, $\log_{10}{(M_{\rm stellar}/M_{\odot})} = 10.67^{+0.15}_{-0.15}$; $z_{\rm HD} = 0.0129$, $\log_{10}{(M_{\rm stellar}/M_{\odot})} = 10.65^{+0.14}_{-0.14}$.~The default aperture was used for SN `2002ef' and SN `1999dg,' while the `small aperture' was used for SN `2001fh.'~$z_{\rm HD}$ denotes the peculiar-velocity corrected Hubble Diagram redshift, although the heliocentric redshift $z_{\rm helio}$ is used to visualize the aperture (the difference between $z_{\rm HD}$ and $z_{\rm helio}$  is negligible except at the lowest redshifts). SN `2002ef' and SN `1999dg' are found to be in high-mass galaxies in this analysis while \Pan~assigned them to the low-mass bin.} 
    \label{fig:inspection}
\end{figure*}

\begin{figure*}
    \centering
    \includegraphics[width=0.45\linewidth]{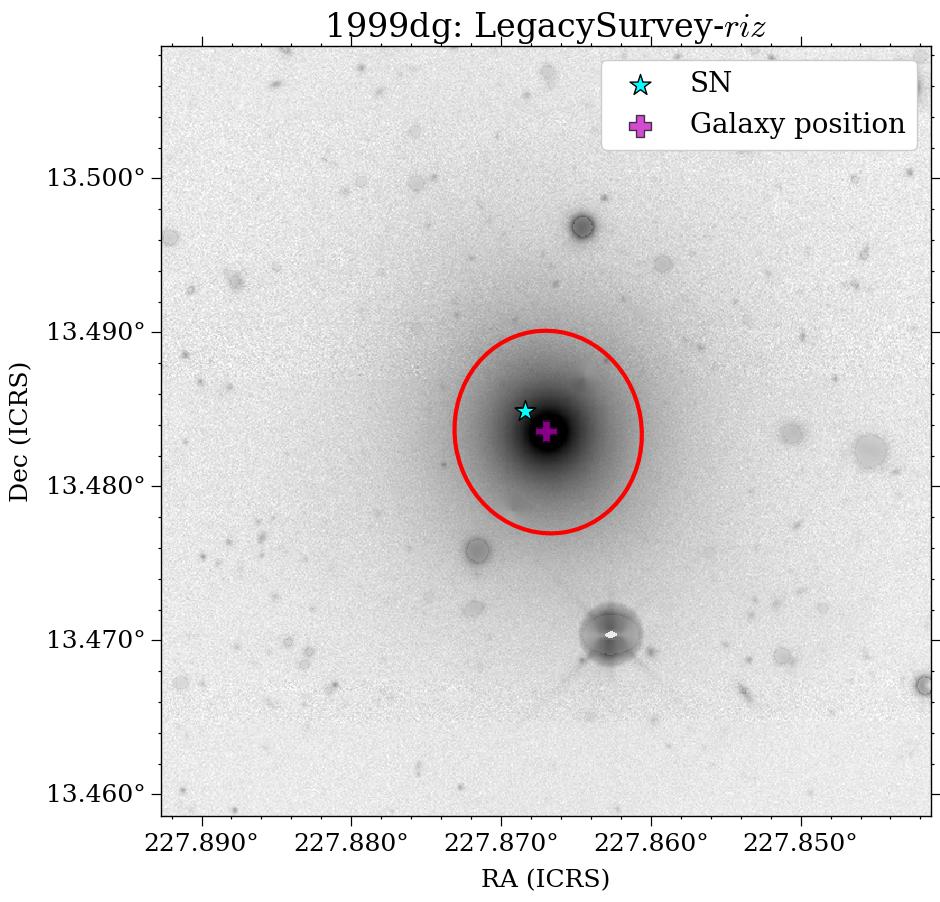}\includegraphics[width=0.45\linewidth]{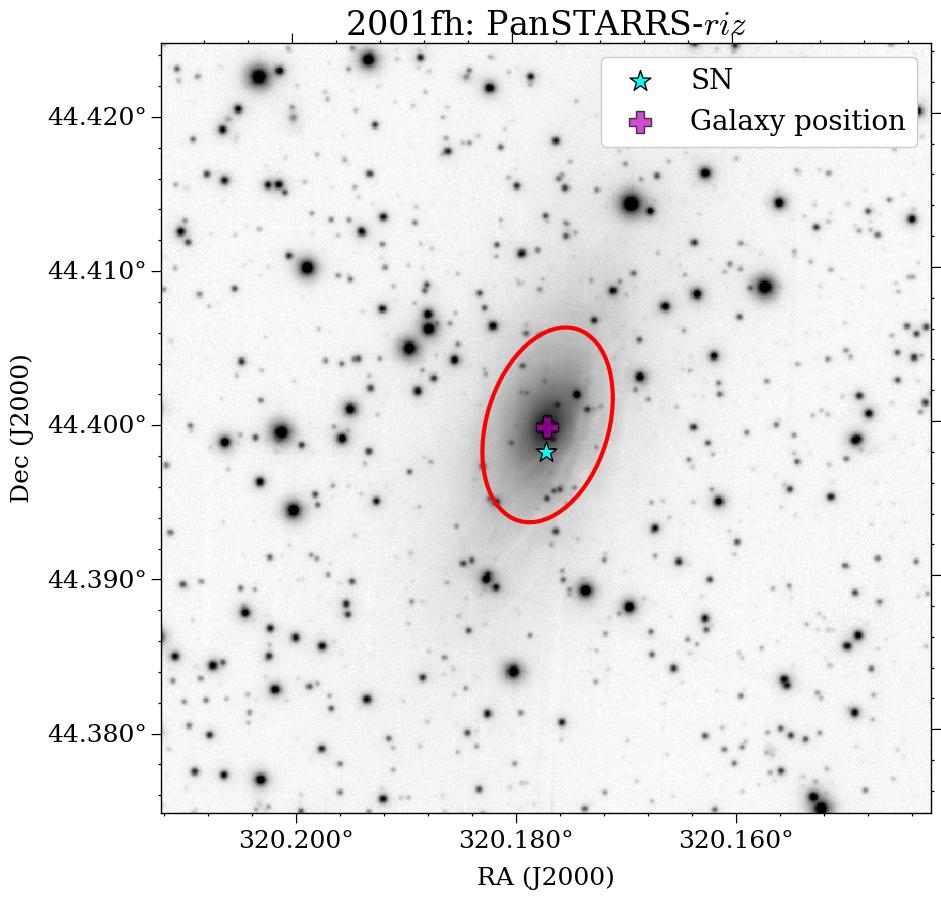}
    \caption{`Small aperture' and `no masking tiny aperture' for SN `1999dg' (Left) and SN `2001fh' (Right).~For both host-galaxies, the selected aperture is now too small to include the outskirts of the galaxies.~The masses using these smaller apertures are: 
    $\log_{10}{(M_{\rm stellar}/M_{\odot})} = 10.63^{+0.12}_{-0.12}$ and $\log_{10}{(M_{\rm stellar}/M_{\odot})} = 10.55^{+0.13}_{-0.13}$ respectively.~This demonstrates that although the aperture selections for the two host-galaxies are not optimal in Figure~\ref{fig:inspection}, the contamination or issues in masking are not enough for the host-galaxies to switch mass-step bins.}
    \label{fig:inspection_details}
\end{figure*}

One major difference from previous analyses is that we \textit{visually inspect} each coadded image for all $~\sim 5000$ (pre-selection cuts) host-galaxies to ensure that the aperture has been appropriately selected.~Based on the inspection, we assign `success scores' on the selected aperture as denoted in Table~\ref{tab:success_scores}.~An ideally selected aperture (success score: 1) would (i) have little or no contamination remaining after masking within the \verb|HostPhot|-computed aperture and (ii) not be a catastrophic failure where incorrect masking of host components or image quality prevents \verb|HostPhot| from selecting an appropriate aperture.~We only retain measurements when the success score is at least 0.5, determined after rerunning with different variations of \verb|HostPhot| parameters - default, small aperture (\verb|eps| = 0.003), large aperture (\verb|eps| = 0.0005), no masking, and no masking tiny aperture (\verb|eps| = 0.01), then selecting the most appropriate option.~For all measurements where the success score is less than 1, we compute the photometry of the variations and the impact on the host-galaxy mass measurements, and hence cosmology in Sections~\ref{sec:results_mass} and \ref{sec:results_cosmo}.~Examples of galaxies with their assigned success scores are shown in Figure~\ref{fig:inspection}.~Figure~\ref{fig:inspection_details} further illustrates that the contaminants and masking issues within the selected apertures for success scores 0.7 and 0.5 do not result in a change in mass-step bins, where the mass-step location is assumed to be $10^{10} M_{\odot}$.

We find that the visual inspection is quite helpful for the low-$z$, $z_{\rm helio} \le 0.10$ host-galaxies, as there can be significant contamination from foreground stars and sources.~For $z_{\rm helio} > 0.10$ host-galaxies, we do not find visual inspection crucial, with less than $2\%$ of hosts being assigned a success score less than 0.5 in the initial run for the DES coadds, although there are some cases where inspection is beneficial, such as when a galaxy previously assumed to be one is a blend of two or more galaxies. 

Once the aperture is determined, \verb|HostPhot| measures the host-galaxy photometry in AB magnitudes for the optical and near-UV, and in Vega magnitudes for near-IR, as is the standard.


\subsection{Spectral Energy Distribution (SED) Fitting} \label{sec:SED_fitting}

Once we have measured the photometry, we use the CIGALE \citep{boquien2019cigale} package to determine host-galaxy properties.~As described in \citet{qu2024host-mismatch}, CIGALE uses the host-galaxy photometry and redshifts to find the best-fit combination of user-input model parameters.~We use the \verb|sfhdelayed|, \verb|m2005|, and the \verb|dustatt_modified_starburst| modules.~In \verb|sfhdelayed|, the delayed star formation history (SFH) is modeled as: 

\begin{equation}
    \mathrm{SFR}(t) \propto \frac{t}{\tau^2} \times \exp{-t/\tau}, \quad 0 \le t \le t_0 ,
\end{equation}

\begin{table}[]
    \centering
    \begin{tabular}{ll}
    \toprule
            Variant &  Explanation \\
    \midrule
    small aperture (success $<$ 1)     &    \verb|eps| = 0.003\\
    large aperture (success $<$ 1)   &     \verb|eps| = 0.0005\\
    no masking (success $<$ 1)  &  Contaminants not masked   \\
    no masking tiny aperture \\ (success $<$ 1)  &  No masking, \verb|eps| = 0.01\\
    optical only  & Optical photometry only \\
    with NIR    & with \textit{2MASS} photometry \\
    
    \verb|m2005| Salpeter IMF  & \verb|m2005| SSP, Salpeter IMF \\
    \verb|bc03| Chabrier IMF   & \verb|bc03| SSP, Chabrier IMF  \\
    \verb|bc03| Salpeter IMF & \verb|bc03| SSP, Salpeter IMF \\
    \bottomrule
    \end{tabular}    
    \caption{List of variants considered in Sections~\ref{sec:results_mass} and~\ref{sec:results_cosmo}. The [bc03] SSP is based on \citet{bc2003ssp}.}
    \label{tab:analysis_var}
\end{table}

\noindent where $t$ denotes the time elapsed since the onset of star formation, $t_0$ is the age of the main stellar population at the epoch of observation, and $\tau$ is the time when the star formation rate (SFR) peaks. Dust attenuation in \verb|dustatt_modified_starburst| is considered based on the starburst attenuation curve of \citet{calzetti2000dust}, extended with the \citet{leitherer2002} curve.

While \citet{qu2024host-mismatch} use the \verb|bc03| library of single stellar populations (SSPs) with a Salpeter initial mass function (IMF) \citep{salpeter1955}, we use the \verb|m2005| \citep{maraston2005} SSP in order to assume a Kroupa IMF \citep{kroupa2001} for our default masses. (CIGALE does not allow the user to assume a Kroupa IMF with \verb|bc03|.)~\citet{smith2020hosts} and \citet{wiseman2020_DES_COADD} also assume a Kroupa IMF but use underlying spectral synthesis templates from \verb|PÉGASE.2|  \citep{PEGASE1997,le_borgne2002_ZPEG} when determining the \Des~host-galaxy masses.~The main SFH parameter grids span: \verb|tau_main| = 1 to 8000 Myr, \verb|age_main| = 1000 to 10000 Myr, and \verb|f_burst| = 0.001 to 0.60, with the \verb|metallicity| = 0.02 and \verb|separation_age| = 10 Myr.~The default CIGALE input parameters used in this analysis will be included in the SN-Unite data-release.~With this configuration, one CIGALE run using 128 cores takes about 2 minutes for 1,000 hosts.~While it is possible to define a finer grid of parameters, increasing the total grid size by a factor of 50 leads to only a small change in the SED fitting results while increasing the runtime to $\sim 1$ hour.~Given our relatively restricted wavelength coverage, the photometry limits our ability to robustly constrain complex star-formation histories.~We therefore checked that  setting \verb|f_burst| = 0 does not affect the host-galaxy stellar mass measurements significantly enough for cosmological constraints to be altered more than for the variants considered in Table~\ref{tab:analysis_var}. 

Additionally, we compare each of the multi-band host-galaxy fluxes from the CIGALE SED fit with the input flux from \verb|HostPhot| to identify \textit{individual} bands with issues. For $ < 3$\% of the host-galaxies, the image cutouts for one or two specific band(s) have issues such as bad pixels, in which case we remove the specific band(s) and rerun the SED fitting. 

\subsubsection{Host-Galaxy Redshifts}

CIGALE requires either a redshift or an explicit distance to set the luminosity distance, which determines the conversion between intrinsic model luminosities and the observed fluxes used in the SED fit.~For the CIGALE redshift (or distance) inputs, we mostly use host-galaxy redshifts from the \Des~and \Pan~data-release, which are all spectroscopic redshifts.~For $z_{\rm HD} \ge 0.01$ hosts, we use the Hubble Diagram redshift, which includes peculiar-velocity corrections.~For hosts at $z_{\rm HD} < 0.01$, we instead use the CMB frame redshift as peculiar-velocity corrections become increasingly uncertain at very low redshifts.~At $z_{\rm CMB} < 0.003$, peculiar velocities become comparable to the Hubble expansion velocity, meaning that redshift-based distances are no longer reliable. Therefore, for the 6 unique hosts where $z_{\rm CMB} < 0.003$ (NGC 4424, M101, M82, NGC 5253, NGC 4526, and NGC 2841), we use direct distance measurements in Mpc from \citet{M82_Distance_sakai1999,NGC2841_Distance_macri2001,NGC4424_NGC4256_Distance_hatt2018,NGC5253_Distance_sabbi2018,M101_Distance_beaton2019}.

\subsection{Host-Galaxy Stellar Mass Uncertainties} \label{sec:mass_uncertainties}

~We use the \verb|bayes.stellar.m_star_total|~and \verb|bayes.stellar.m_star_total_err| to compute our log stellar mass and log stellar mass error estimates.~We set \verb|additionalerror|, which is the relative error added in quadrature for fluxes and the extensive properties, to the CIGALE default value of 0.1.~While this choice is conservative in the sense that the reduced $\chi^2$ of the SED fits are driven to well below unity, we note that this accounts for various systematic uncertainties arising from modeling choices (e.g., using photometry from multiple different surveys, choosing a particular IMF or a particular SSP library as detailed in Section~\ref{sec:analysis_var}). The resulting median log host-galaxy stellar mass uncertainty for the SN-Unite analysis is 0.23 dex.~For more details on the host-galaxy stellar mass uncertainties, see Appendix~\ref{sec:appendixA}. The various host-galaxy log stellar mass uncertainty distributions are shown in Figure~\ref{fig:log_mass_err_distributions}.

\subsection{Analysis Variants} \label{sec:analysis_var}

Although the host-galaxy stellar mass is one of the most straightforward host-galaxy properties to measure, various analysis choices can lead to different mass measurements even for the same host.~Table~\ref{tab:analysis_var} lists the different analysis variants we consider in Sections~\ref{sec:results_mass} and~\ref{sec:results_cosmo} to assess whether our conclusions are sensitive to the analysis choices. 

\subsection{Unite Sample Variants and Data-Release Masses} \label{sec:methodology_dr_masses}

As advertised in Section~\ref{sec:intro}, the newly derived SN-Unite host-galaxy stellar mass measurements and the resulting cosmological constraints are compared to the \Des~and \Pan~ data-release measurements and cosmological constraints in Sections~\ref{sec:results_mass_with_previous} and~\ref{sec:results_cosmo_DR_masses}. 



In Section~\ref{sec:results_cosmo_DR_masses}, we assess the impact on cosmological constraints of replacing the \Pan~and \Des~data-release (DR) masses with the newly derived SN-Unite masses for each of these three sample variants:

\begin{enumerate}
    \item \textbf{The Unite sample} - for the original data-release (DR) masses (as opposed to newly derived masses in this work), we use the \Pan~DR masses for the low-$z$ hosts common to both \Des~and \Pan. 
    \item \textbf{The Unite-\Des~subsample} - for the data-release (DR) masses, we use the \Des~DR masses throughout all redshifts including for the external low-$z$ datasets.~The Unite-\Des~DR subsample is similar to the DES-Dovekie sample, with some exceptions, described in \citet{SN-Unite}.~Notably, SN-Unite applies a $P_{\rm Ia} >0.8$ cut, resulting in fewer SNe Ia, and low-$z$ SNe Ia with multiple observations are treated differently. 
    \item \textbf{The Unite-\Pan~subsample} - for the data-release (DR) masses, we use the \Pan~data-release masses throughout all redshifts.~Unlike the Unite-\Des~subsample, we include all of the DES-SN3YR sample including those without host spectroscopic redshifts, and perform bias corrections using the DES-SN3YR methodology.~Note that the Unite-\Pan~subsample considered in this work therefore contains more SNe Ia than the \Pan~SNe Ia \textit{common} to Unite discussed in Section 9.2.2 of \citet{SN-Unite}.~The Unite-\Pan~subsample, however, is \textit{not} identical to the official \Pan~sample both because of stricter selection cuts imposed in SN-Unite, and numerous updates in the analysis pipeline since the \Pan~analysis.
\end{enumerate}


\subsection{Quantifying the Impact on Cosmology}

\begin{figure*}
    \centering
    \captionsetup{skip=0pt}
    \includegraphics[width=0.95\linewidth]{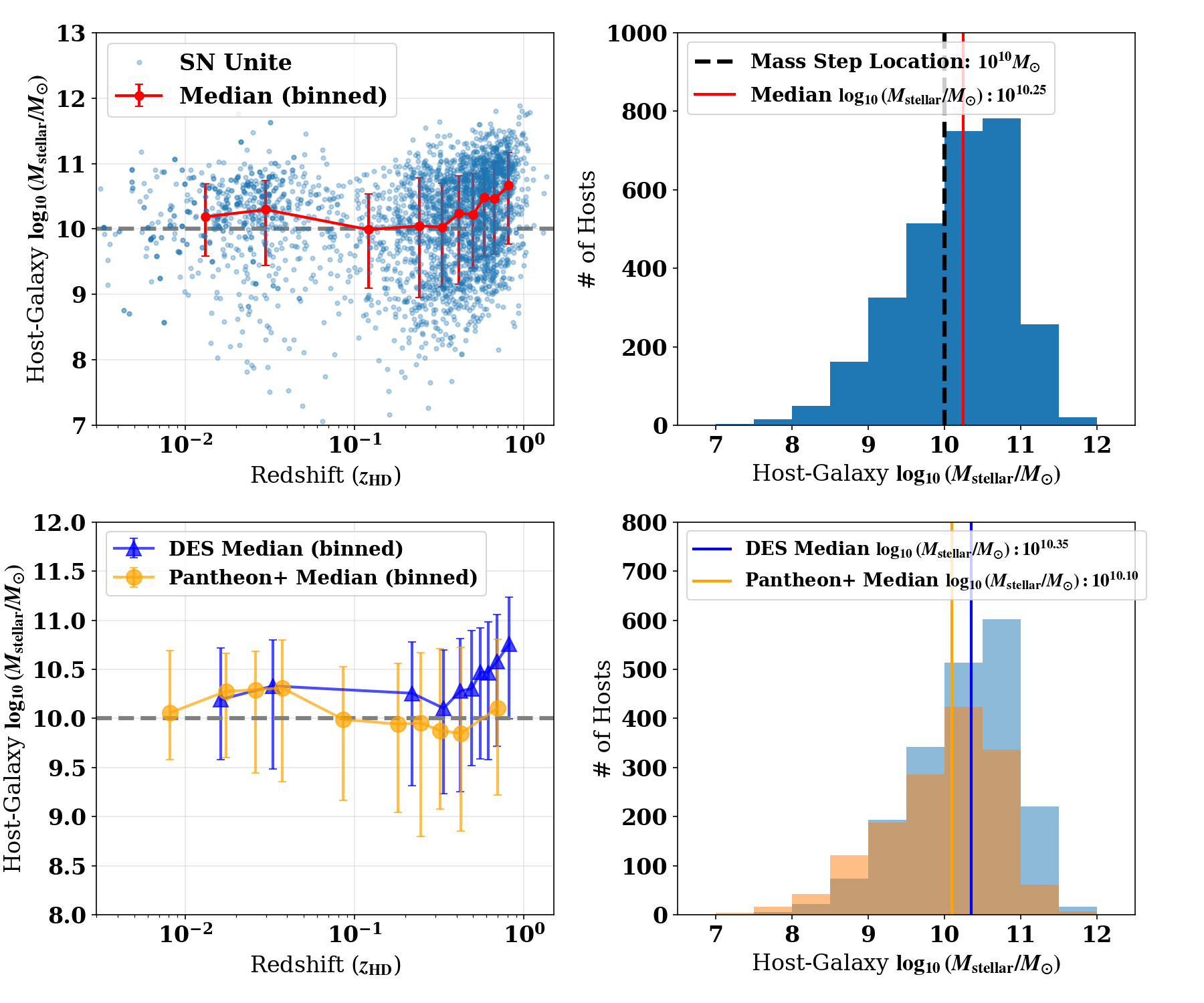}
    \caption{(Top left) SN-Unite default host-galaxy log stellar masses vs. redshift.~The mass measurements are split into 10 equal size bins in redshift, with the median and the 16th and 84th percentiles shown within each bin.~(Top right) SN-Unite host-galaxy log mass distribution with the mass-step location and the median log mass shown.~About 63\% of our sample is above the mass-step.~(Bottom left and right) Same as above but shown for \Des~and \Pan~samples within SN-Unite separately. Host mass uncertainties are not shown for clarity. }
    \label{fig:Unite_masses_default}
\end{figure*}

As presented in \citet{SN-Unite}, we use the BEAMS with Bias Corrections framework \citep[BBC,][]{2017_BiasCor} implemented in \verb|SNANA|\footnote{https://github.com/RickKessler/SNANA} \citep[SuperNova ANAlysis,][]{kessler2009snana} integrated within \verb|PIPPIN| \citep{hinton2020pippin} to obtain our bias corrected Hubble diagrams.~We treat each of the variations considered in Table~\ref{tab:analysis_var} as a systematic to assess its impact on cosmology, as described in Sections 4.9 and 6 of \citet{SN-Unite}.~This allows us to compute the shift of cosmological parameters due to each systematic from the baseline `no-systematics' results, which include statistical uncertainties only, in Flat$w$CDM. For the subsample and data-release (DR) masses analyses, we rerun \verb|PIPPIN| separately, including all the nominal systematics as in the main SN-Unite sample presented in \citet{SN-Unite}, with the exception of contamination-related systematics for the spectroscopic Unite-\Pan~subsample. 

To obtain cosmological constraints for the subsample and data-release masses comparison, including combinations with BAO and the CMB, we use the nested sampler \texttt{Nautilus}\footnote{\url{https://github.com/johannesulf/nautilus}} \citep{lange2023_nautilus} within the \texttt{CosmoSIS} \citep{ZUNTZ201545} framework, same as the main SN-Unite sample, highlighted in Section 7 of \citet{SN-Unite}. The cosmological constraints shown in this work do not contain weak lensing corrections described in Section 4.7 of \citet{SN-Unite}, in order to compare the subsample results to \Pan~and \Des~data-release constraints (which did not include weak lensing corrections) as directly as possible. 


All of the analysis pipelines presented in this work including the host-galaxy stellar mass measurements were frozen before unblinding SN-Unite (see Section 5.1 of \citet{SN-Unite}). 

\subsubsection{Impact of Host-galaxy mass uncertainties on Cosmology}

Unlike in \Des~and DES-Dovekie (or \Pan), we incorporate our host-galaxy log stellar mass uncertainties into the systematic error budget.~We draw 10 realizations of the SN-Unite host-galaxy stellar masses from $\mathcal{N}(\log_{10} M_{\rm stellar}, \ln (10)\frac{\sigma_{M_{\rm stellar}}}{M_{\rm stellar} })$ where $M_{\rm stellar}$ and $\sigma_{M_{\rm {stellar}}}$ are \verb|bayes.stellar.m_star_total| and \verb|bayes.stellar.m_star_total_err| from CIGALE respectively.~As discussed in Section 4.10 in \citet{SN-Unite}, each of the realizations is treated as an analysis variant (with a weight of 1/10 each) in SN-Unite, which allows us to estimate the level of contribution from host-galaxy stellar mass uncertainties to the systematic uncertainties in cosmological parameters for SN-Unite.  


\section{Mass Measurements} \label{sec:results_mass}

In this Section, we present our new mass measurements for the SN-Unite cosmological analysis, discuss their sensitivity to photometry and SED fitting choices as well as compare them to previous measurements.

\begin{table*}
\centering
\begin{tabular}{clccc}
\hline
Row & Variant & $\langle \Delta \log_{10} (M_{\rm stellar}/M_{\odot}) \rangle$ (rms) & High $\rightarrow$ Low & Low $\rightarrow$ High \\
\hline
1 & Fiducial (SN Unite default) & -- & -- & -- \\

2 & Small aperture & $-0.004 \pm 0.003$ (0.182) & 26 (0.9\%) & 18 (0.6\%) \\
3 & Large aperture & $0.027 \pm 0.004$ (0.198) & 1 (0.0\%) & 24 (0.8\%) \\
4 & No masking & $0.053 \pm 0.005$ (0.257) & 0 (0.0\%) & 53 (1.9\%) \\
5 & No masking tiny aperture & $-0.019 \pm 0.004$ (0.204) & 51 (1.8\%) & 21 (0.7\%) \\
6 & Optical only & $-0.020 \pm 0.002$ (0.093) & 26 (0.9\%) & 29 (1.0\%) \\
7 & With NIR & $-0.013 \pm 0.002$ (0.110) & 35 (1.2\%) & 21 (0.7\%) \\
8 & Salpeter IMF & $0.198 \pm 0.000$ (0.014) & 0 (0.0\%) & 218 (7.7\%) \\
9 & bc03 Chabrier IMF & $0.053 \pm 0.001$ (0.046) & 2 (0.1\%) & 63 (2.2\%) \\
10 & bc03 Salpeter IMF & $0.299 \pm 0.001$ (0.046) & 0 (0.0\%) & 328 (11.5\%) \\

\hline
\end{tabular}
\caption{
Comparison of host-galaxy stellar masses derived using different assumptions relative to the default SN-Unite measurements.
$\Delta \log_{10} (M_{\rm stellar}/M_{\odot}) = \log_{10} (M_{\rm stellar}^{\mathrm{variant}}/M_{\odot}) - \log_{10} (M_{\rm stellar}^{\mathrm{SN-Unite\ default}}/M_{\odot})
$, and we show the mean, error on the mean, and the rms of $\Delta \log_{10} (M_{\rm stellar}/M_{\odot})$. 
High $\rightarrow$ Low indicates hosts with
$\log_{10} (M_{\rm stellar}^{\mathrm{SN-Unite\ default}}/M_{\odot})>10$ and
$\log_{10} (M_{\rm stellar}^{\mathrm{variant}}/M_{\odot})<10$, and vice versa.
}
\label{tab:unite_mass_variants}
\end{table*}

\subsection{Nominal Results}
\label{sec:results_mass_nominal}

In Figure~\ref{fig:Unite_masses_default}, we present the nominal host-galaxy log stellar masses used in the SN-Unite analysis.~In the left plot, we show the log mass against $z_{\rm HD}$, with the median, 16th, and 84th percentiles of the 10 equally sized bins (in redshift) overplotted.~We use the median rather than the mean to mitigate the impact of outliers.~On the right plot, we show the distribution of the host log mass.~The median log mass is found to be $\log_{10} (M_{\rm stellar}/M_{\odot}) = 10.25 $, with about 63\% of the sample being above the mass-step location of $10^{10} M_{\odot}$. For reference, in \Des, about 68\% of the galaxies were found to be high-mass \citep{vincenzi2024DES_systematics}, while for \Pan, about 54\% of the host-galaxies are above the mass step. This reflects the differences in the selection functions between \Des~and \Pan, especially at the higher redshifts, as shown in the bottom left of Figure~\ref{fig:Unite_masses_default}.  

\begin{figure*}
    \centering
    \captionsetup{skip=0pt}
    \includegraphics[width=0.95\linewidth]{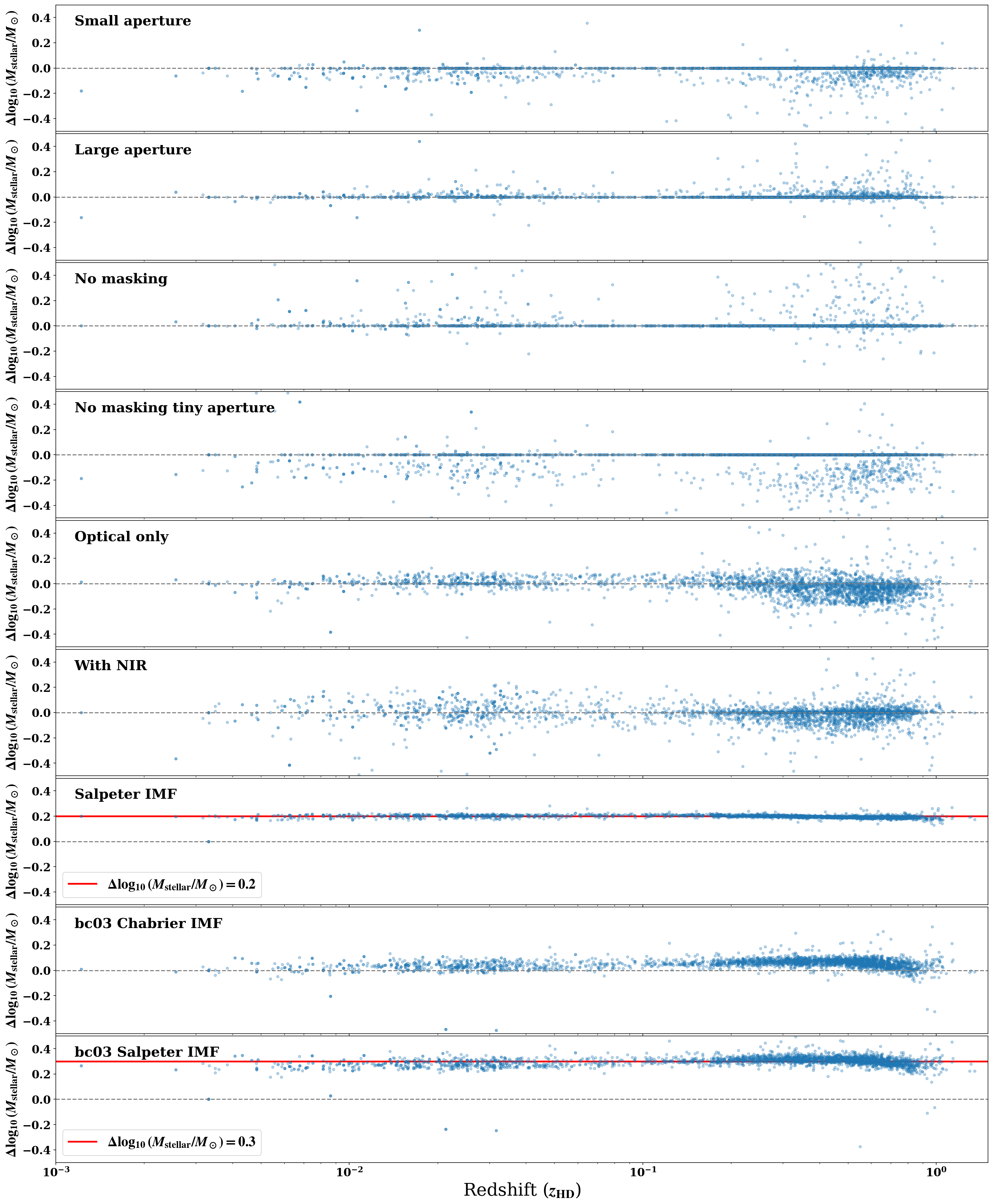}
    \caption{$\Delta \log_{10} (M_{\rm stellar}/M_{\odot}) = \log_{10} (M_{\rm stellar}^{\mathrm{variant}}/M_{\odot}) - \log_{10} (M_{\rm stellar}^{\mathrm{SN-Unite\ default}}/M_{\odot})
$ vs. $z_{\rm HD}$ between the variants introduced in Section~\ref{sec:analysis_var} and the default SN-Unite mass measurements. For the Salpeter IMF variants, we also show the mean of $\Delta \log_{10} (M_{\rm stellar}/M_{\odot})$ for reference.}
    \label{fig:Unite_vs_Mass_variants}
\end{figure*}

Figure~\ref{fig:Unite_vs_Mass_variants} shows the difference $\Delta \log_{10} (M_{\rm stellar}/M_{\odot}) = \log_{10} (M_{\rm stellar}^{\mathrm{variant}}/M_{\odot}) - \log_{10} (M_{\rm stellar}^{\mathrm{SN-Unite\ default}}/M_{\odot})
$ between the variants introduced in Section~\ref{sec:analysis_var} and the default SN-Unite mass measurements against redshift. We see a negligible trend with redshift except when we limit the photometry to optical only, or include the NIR along with the default optical and NUV. We find in Section~\ref{sec:results_cosmo_ana_choice} that the slight redshift-dependent trend that varies depending on the wavelength coverage has a negligible impact on cosmological constraints.  

Table~\ref{tab:unite_mass_variants} shows the mean, error on the mean, and the rms of the  difference $\Delta \log_{10} (M_{\rm stellar}/M_{\odot})
$ between the variants introduced in Section~\ref{sec:analysis_var} and the default SN-Unite mass measurements.~Assuming a Salpeter IMF gives $0.2$ to $0.3$ dex higher log mass measurements (see the Salpeter IMF panels of Figure~\ref{fig:Unite_vs_Mass_variants}), which is expected as the Salpeter IMF predicts a much higher fraction of low-mass stars than more recently devised IMFs. Other variants are consistent with the default measurements, both in terms of the difference as well as the number of hosts that move across the mass-step.~The latter can be particularly important given that the mass-step is modeled as a step function at $10^{10} M_{\odot}$, meaning that if we use the \verb|bc03| Salpeter IMF, more than 10\% of the SNe Ia will have distance moduli more than $0.05$ mag brighter than the default case.~In Section~\ref{sec:results_cosmo}, we assess the offset in cosmological parameters when we use the variants instead of the default mass measurements.

\subsection{Comparison with previous measurements}
\label{sec:results_mass_with_previous}

Given that the differences between the overlapping low-$z$ mass-measurements in \Des~and \Pan~impact cosmological constraints on time-evolving dark energy \citep{vincenzi2025response}, it is important to compare our new measurements with host-galaxy stellar mass measurements from both the \Des~and \Pan~data-release (DR).

\begin{figure*}
    \centering
    \captionsetup{skip=0pt}
    \includegraphics[width=0.95\linewidth]{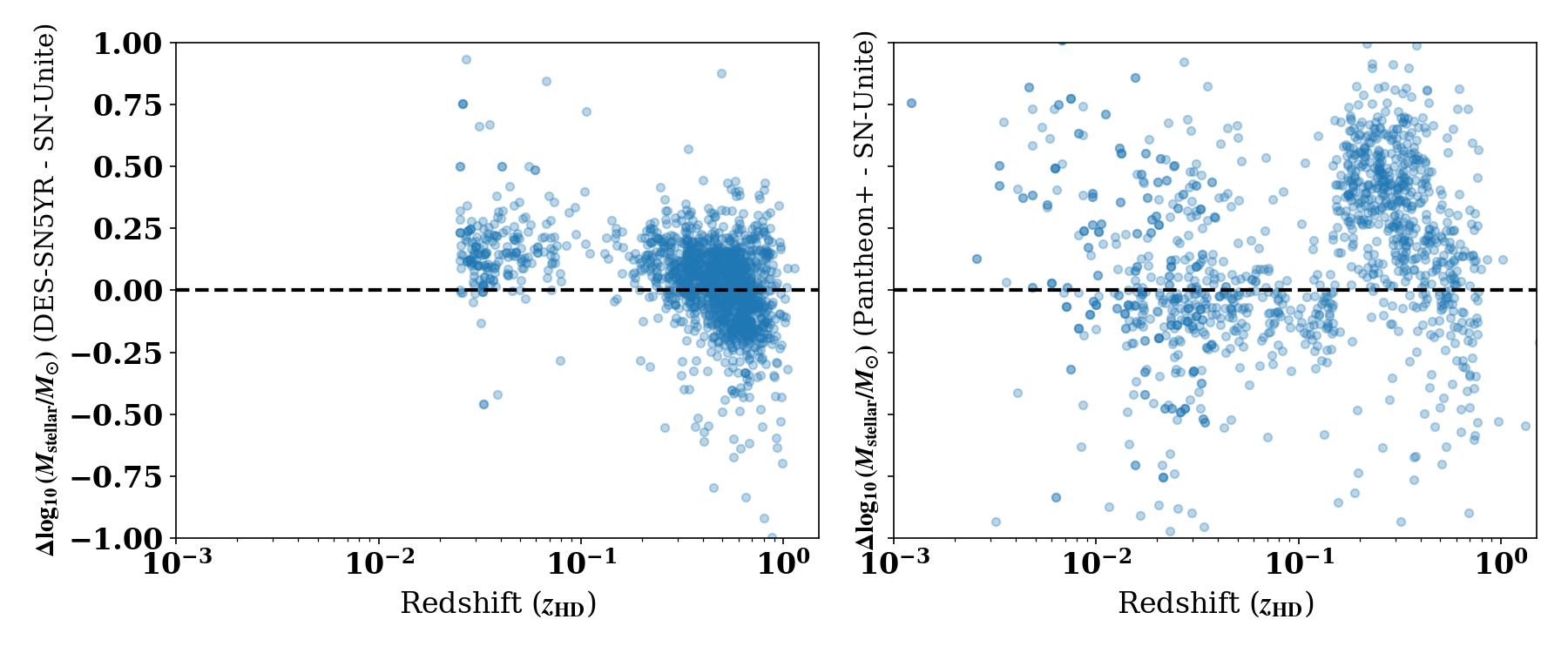}
    \caption{Difference between SN-Unite and \Des~host-galaxy log stellar masses (left) and SN-Unite and \Pan~(right), used for the SN-Unite cosmological analysis.~While \Des~measurements are largely consistent with the new SN-Unite measurements, \Pan~masses are noticeably discrepant in a redshift dependent way.~At $z_{\rm HD} < 0.15$, \Pan~masses tend to be lower than SN-Unite, while at $ z_{\rm HD} \ge 0.15$, \Pan~masses tend to be higher. This is in part because \Pan~assumed the Chabrier IMF when re-deriving the $z_{\rm HD} < 0.15$ SN Ia host-galaxy masses from historical samples, resulting in a discrepancy with the historical host masses they adopted for $z_{\rm HD} > 0.15$. Uncertainties are not shown for clarity, but of order 0.2 dex.}
    \label{fig:Unite_vs_DR_masses}
\end{figure*}

\begin{table*}
\centering
\begin{tabular}{lccccc}
\hline
Sample & $z$ bin &
$\langle \Delta \log_{10} (M_{\rm stellar}/M_{\odot}) \rangle$ (rms) &
$N_{\mathrm{used}}/N_{\mathrm{tot}}$ &
High $\rightarrow$ Low &
Low $\rightarrow$ High \\
\hline

DES-SN5YR & Low-$z$ ($z_{\rm HD}<0.10$) & $0.160 \pm 0.009$ (0.114) & 171/184 & 1 (0.6\%) & 11 (6.4\%) \\
DES-SN5YR & High-$z$ ($z_{\rm HD}\ge 0.10$) & $0.017 \pm 0.004$ (0.162) & 1357/1373 & 17 (1.3\%) & 33 (2.4\%) \\
Pantheon+ & Low-$z$ ($z_{\rm HD}<0.15$) & $-0.028 \pm 0.019$ (0.471) & 643/721 & 100 (15.6\%) & 39 (6.1\%) \\
Pantheon+ & High-$z$ ($z_{\rm HD}\ge 0.15$) & $0.283 \pm 0.015$ (0.385) & 636/666 & 20 (3.1\%) & 75 (11.8\%) \\

\hline
\end{tabular}
\caption{
Comparison of host-galaxy stellar masses between external samples and SN-Unite.~Differences are defined as
$\Delta \log_{10} (M_{\rm stellar}/M_{\odot}) = \log_{10} (M^{\rm Previous}_{\rm stellar}/M_{\odot}) - \log_{10} (M^{\rm SN-Unite}_{\rm stellar}/M_{\odot})$, and we show the mean, error on the mean, and the rms of $\Delta \log_{10} (M_{\rm stellar}/M_{\odot})$. Due to non-negligible catastrophic outliers especially for \Pan, 
we quote $\langle \Delta \log_{10} (M_{\rm stellar}/M_{\odot}) \rangle$ (rms) values computed after imposing a 
$5\sigma$ clip based on the median absolute deviation of $\Delta \log_{10} (M_{\rm stellar}/M_{\odot})$. High $\rightarrow$ Low indicates hosts with
$\log_{10} (M^{\rm SN-Unite}_{\rm stellar}/M_{\odot})>10$ and
$\log_{10} (M^{\rm Previous}_{\rm stellar}/M_{\odot})<10$, and vice versa; this includes all hosts in the \Pan~data-release Hubble diagram. 
Low-$z$ and high-$z$ selections follow the definitions adopted for
DES (split at $z_{\rm HD} = 0.10$) and \Pan~($z_{\rm HD} = 0.15$) respectively.
}
\label{tab:unite_vs_previous_mass}

\end{table*}

Figure~\ref{fig:Unite_vs_DR_masses} shows the difference between the \Des~and SN-Unite host-galaxy log stellar masses (left) and SN-Unite and \Pan~host-galaxy log stellar masses (right).~We find that the newly derived SN-Unite masses are largely consistent with the \Des~measurements throughout all redshifts, with the median SN-Unite log mass being slightly lower than the median \Des~log mass by about 0.16 dex at $z_{\rm HD} < 0.10$ and comparable to \Des~at $z_{\rm HD} \ge 0.10$.~Assuming an \verb|additionalerror| = 0.0 instead of 0.1 and therefore removing systematic uncertainties (analogous to \Des, which only includes statistical uncertainties in host stellar mass estimates) in SN-Unite masses reduces the scatter in $\Delta \log_{10} (M_{\rm stellar}/M_{\odot})$ at  $z_{\rm HD} \ge 0.10$ and $\Delta \log_{10} (M_{\rm stellar}/M_{\odot}) \approx 0.1$ dex throughout all redshifts.~This remaining slight offset likely arises from different analysis choices in the SED fitting.~In \Des, the SED fitting was done using the underlying libraries from the \verb|PÉGASE.2| spectral synthesis templates \citep{PEGASE1997,le_borgne2002_ZPEG}.~\citet{smith2020hosts} found that using the \citet{maraston2005} template results in a systematic offset of $0.11 \pm 0.01 $ dex for the DES-SN3YR survey, with the \verb|PÉGASE.2| masses being more massive, which is consistent with our finding. We note that the SN-Unite \textit{griz} host-galaxy photometry for the \Des~sample is quite consistent with those of \Des~\citep{wiseman2020_DES_COADD,sanchez2024DES_DR} as reported in Table~\ref{tab:w20_coadd_unite_photometry}, with the median values agreeing to better than 0.015 mag, suggesting that the contribution from photometry differences is minor.

~In contrast, we see considerable discrepancy between the SN-Unite and \Pan~masses, with SN-Unite masses being generally higher than \Pan~at $z_{\rm HD} < 0.15$ and \Pan~masses being considerably higher at $z_{\rm HD} > 0.15$.~This redshift-dependent discrepancy in the \Pan~host-galaxy masses is also seen in \citet{hoyt2026union3.1_hosts}.~Part of this discrepancy can be attributed to the \Pan~team re-deriving host-galaxy stellar masses from historical samples, but only for $z_{\rm HD} < 0.15$ using a Chabrier IMF \citep{chabrier2003}, while the historical mass measurements at $z_{\rm HD}> 0.15$ are typically more consistent with SED fitting assuming a Salpeter IMF \citep{sullivan2010_mass-step,Betoule2014JLA,sako2018SDSS}. In Table~\ref{tab:unite_vs_previous_mass}, we show the mean offset, the error on the mean, as well as the rms between the SN-Unite default measurements and \Des~/ \Pan~split into high-$z$ / low-$z$.~We find that the mean mass offset between assuming a Salpeter IMF and a Chabrier IMF is around 0.25 dex in Table~\ref{tab:unite_mass_variants}, and up to 60\% of the discrepancy between $z_{\rm HD}> 0.15$ and $z_{\rm HD} < 0.15$ host-galaxy log masses in \Pan~can be explained by their choice of IMF at $z_{\rm HD} < 0.15$.~Additionally, \Pan~assigned hosts where they could not obtain host masses to the low-mass bin, or $\log_{10} (M_{\rm stellar}/M_{\odot}) = 7 $ after confirming that the hosts are faint and have not been mis-identified \citep{scolnic2022pantheon+}. In SN-Unite, we are able to determine host-galaxy stellar masses for 80 hosts across all redshifts in the SN-Unite sample where \Pan~was not able to determine the mass, likely due to the deeper images used for this analysis. We also find that for 43 of these hosts, $\log_{10} (M_{\rm stellar}/M_{\odot}) \ge 10 $. This is similar to the findings of \citet{hoyt2026union3.1_hosts}.


\subsubsection{Which Initial Mass Function?}
\label{sec:IMF}

\begin{figure}
    \centering
    \includegraphics[width=0.95\linewidth]{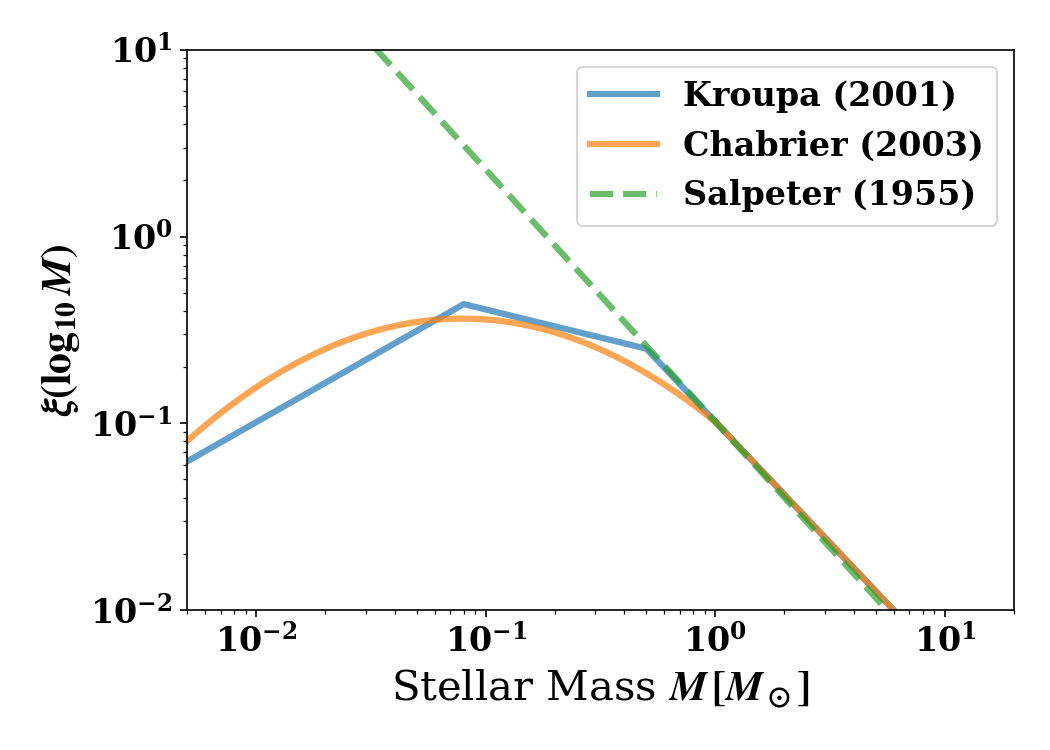}
    \caption{The Kroupa, Chabrier and Salpeter initial mass functions for individual stars.~The high-mass tail is essentially the same for the three while the Kroupa and Chabrier IMFs predict fewer low-mass stars.~The curves have been normalized to the same value at $M = 1 M_{\odot}$.}
    \label{fig:IMF}
\end{figure}

Given that host-galaxy mass measurements are most sensitive to the choice of IMF, one may question, `Which IMF should we then choose for SN Ia host-galaxy mass measurements?'~We briefly highlight the difference in the three IMFs. 

The initial mass function (IMF) is defined as the number of stars $N$ in volume $V$ observed at a time $t$ per log mass interval $d\log m$ \citep{salpeter1955,chabrier2003review}: 

\begin{equation}
    \xi(\log_{10} M) = \frac{d(N/V)}{d\log_{10 }M} = \frac{dn}{d\log_{10} M} 
\end{equation}

\noindent with $n = N/V$ being the stellar number density.~In Figure~\ref{fig:IMF}, we show the Kroupa, Chabrier and Salpeter IMFs. The Salpeter IMF is described as a power-law, while the Kroupa IMF is a broken power-law, and the Chabrier IMF is a log-normal at $M < 1 M_{\odot}$.~For explicit expressions for the IMFs, see Appendix~\ref{sec:appendixB}. While general consensus for the past few decades has been that the high-mass tail is well-described by the power-law given by \citet{salpeter1955}, more recent observations found a discrepancy in the low-mass end, hence resulting in the Kroupa and Chabrier IMFs.~There is not a clear preference between the Chabrier IMF and the Kroupa IMF, meaning they are both widely used currently; we note that the difference between the Kroupa and Chabrier IMFs at the low-mass end is quite small. 

\begin{table*}
\centering
\caption{Impact of analysis choices on the SALT3 light-curve parameters as well as cosmology. The variations are treated as systematics and are compared to the baseline with statistical uncertainties only.~$\Delta Q_H$ ($\Delta w$) represents the shift in $Q_H$ ($w$) relative to the `no systematics' baseline while $\Delta \chi^2$ shows the change in fit quality relative to baseline.~Errors on $\alpha$, $\beta$, and $\gamma$ are shown in parentheses.}
\begin{tabular}{l|ccc|cc|ccccc}
\toprule
\multirow{2}{*}{Systematic} & \multicolumn{3}{c|}{SALT3 Parameters} & \multicolumn{2}{c|}{Fit Quality} & \multicolumn{5}{c}{Cosmology} \\
\cmidrule(lr){2-4} \cmidrule(lr){5-6} \cmidrule(lr){7-11}
 & $\alpha$ & $\beta$ & $\gamma$ & $\sigma_{\rm int}$ & RMS & $\Delta \chi^2$ & $\Delta w$ & $\sigma_w$ & $\Delta Q_{H}$ & $\sigma_{Q_{H}}$ \\
\midrule
\addlinespace[0.5em]
baseline (no systematics) & 0.159(3) & 3.08(3) & 0.040(6) & 0.037 & 0.177 & 0.0 & +0.000 & 0.096 & +0.000 & 0.015 \\
\midrule
\addlinespace[0.5em]
mass step loc. 10.2 & 0.160(3) & 3.08(3) & 0.043(6) & 0.036 & 0.177 & -8.3 & +0.009 & 0.095 & -0.004 & 0.015 \\
optical only & 0.160(3) & 3.09(3) & 0.042(6) & 0.037 & 0.177 & 3.9 & +0.002 & 0.096 & -0.000 & 0.015 \\
with near-IR & 0.159(3) & 3.08(3) & 0.038(6) & 0.038 & 0.177 & 13.6 & +0.001 & 0.095 & -0.000 & 0.015 \\
Salpeter & 0.158(3) & 3.11(3) & 0.040(6) & 0.044 & 0.177 & 6.1 & -0.003 & 0.096 & +0.003 & 0.015 \\
bc03 Chabrier & 0.159(3) & 3.09(3) & 0.038(6) & 0.039 & 0.177 & 6.2 & -0.002 & 0.096 & +0.000 & 0.015 \\
bc03 Salpeter (mass-step loc. 10.5) & 0.160(3) & 3.11(3) & 0.050(6) & 0.043 & 0.177 & -18.1 & +0.002 & 0.095 & -0.003 & 0.015 \\
bc03 Salpeter & 0.157(3) & 3.13(3) & 0.039(7) & 0.048 & 0.178 & 17.2 & -0.005 & 0.096 & +0.003 & 0.015 \\
small aperture & 0.158(3) & 3.08(3) & 0.036(6) & 0.038 & 0.177 & 11.7 & +0.000 & 0.096 & -0.000 & 0.015 \\
large aperture & 0.159(3) & 3.09(3) & 0.040(6) & 0.038 & 0.177 & 0.6 & -0.000 & 0.096 & -0.000 & 0.015 \\
no masking & 0.159(3) & 3.08(3) & 0.042(6) & 0.038 & 0.177 & -3.2 & +0.004 & 0.096 & +0.001 & 0.015 \\
no masking tiny aperture & 0.159(3) & 3.07(3) & 0.039(6) & 0.036 & 0.177 & -2.7 & -0.003 & 0.096 & -0.001 & 0.015 \\
host mass uncertainties & 0.159(1) & 3.08(1) & 0.037(2) & 0.038 & 0.178 & 21.2 & -0.009 & 0.096 & -0.004 & 0.015 \\
\bottomrule
\end{tabular}
\label{tab:cosmo_variants}
\end{table*}

As such, we find that the choice of the IMF itself is not very important in determining SN Ia host-galaxy mass measurements for cosmological analysis. Our mass measurements, as well as our cosmological constraints, are quite consistent whether we use the Kroupa or Chabrier IMF (with the \verb|bc03| template). While using the Salpeter IMF systematically increases the mass measurements by $0.2$ dex, we find in Section~\ref{sec:results_cosmo_ana_choice} that the impact on cosmology is small.

In the next section, we assess the impact of using various host-galaxy mass measurements on cosmological parameters.

\section{Impact on Cosmology} 
\label{sec:results_cosmo}

We assess the impact of host-galaxy stellar mass measurements on cosmology in two cases: (i) replacing the default SN-Unite masses with variants including those presented in Section~\ref{sec:results_mass_nominal} and Table~\ref{tab:unite_mass_variants} and (ii) updating the \Des~and \Pan~data release (DR) mass measurements to the new SN-Unite masses determined in this work for SN-Unite.~For case (ii), we examine the Unite-\Des~subsample, the Unite-\Pan~subsample, and the full SN-Unite sample individually.~Case (ii) is done for SN Ia only datasets in Flat$w$CDM, as well as Flat$w_0w_a$CDM when combined with the BAO and CMB. 

As in the main SN-Unite analysis \citep{SN-Unite}, in order to present a single non-degenerate parameter constraint in the $\Omega_m$-$w$ plane, we use the parameter $Q_H(z=0.2)$ (fixing $z = 0.2$ subsequently) defined in Section 3 of \citet{camilleri2024_beyondLCDM} as:  

\begin{equation}
    Q_H(z) = \frac{1}{2}[ \Omega_{\mathrm{m}} a^{-3} + \Omega_{\rm de} (1+3w) a^{-3(1+w)}], 
\end{equation}

\noindent Note that $Q_H \equiv -\ddot{a}/(aH_0^2)\equiv q(H/H_0)^2$ and $a = (1+z)^{-1}$.~$Q_{H}$ plays a similar role for SN Ia analyses as $S_8$ does in weak lensing analyses, breaking the degeneracy with $\Omega_m$ in order to isolate the combination of parameters that optimizes the constraining power of the SN Ia data.~Shifts along the degeneracy direction in the $\Omega_m$-$w$ plane have little impact on the cosmological expansion inferred from SNe Ia, although they can affect the resulting constraints when SNe Ia are combined with other probes that have different degeneracy directions. Shifts in $Q_H$, perpendicular to the degeneracy direction, are more indicative of changes in the expansion history inferred from SNe Ia.~We additionally provide shifts in the equation of state $\Delta w$ for reference. 


\subsection{Sensitivity to Analysis Choices}
\label{sec:results_cosmo_ana_choice}

In Table~\ref{tab:cosmo_variants}, we show the 
shifts in $Q_{H}$, $w$, and $\chi^2$ with respect to the baseline (no systematics) case for SN-Unite, as well as the SALT3 light-curve parameters and their uncertainties for the mass measurement variations (4 photometry and 6 SED fitting choices).~In addition, `mass step loc.~10.2' is when the mass step location is set to $10^{10.2} M_{\odot}$ instead of $10^{10} M_{\odot}$ and `host mass uncertainties' shows the shift in cosmological parameters due to incorporating the SN-Unite host-galaxy stellar mass uncertainties into the cosmological parameters systematic error budget. 


We first see that for all cases, the significance of the mass-step ($\gamma$) is over $6\sigma$, with the baseline being $\gamma = 0.040 \pm 0.006$, slightly higher than the significance reported in \Des~(5$\sigma$) and DES-Dovekie (4$\sigma$). $\Delta Q_{H, \rm syst}$ is less than $0.3\sigma$ (less than 0.1$\sigma$ for $w$) for all mass measurement variations, showing that our cosmological results are robust even with some changes in our mass measurement pipeline.~We note that $\Delta \chi^2$ is -8.3 when we shift the mass-step location to 10.2, and $-18.1$ when the mass-step location is taken to be 10.5 for the \verb|bc03| Salpeter IMF variant (median $ \log_{10} (M_{\rm stellar}/M_{\odot}) = 10.55$).~This suggests that the true mass-step location is slightly higher, around the median host-galaxy stellar mass of the sample.~Ideally, the mass-step location should be fit for each SN Ia sample, but the current \verb|SNANA| pipeline does not provide a robust fitting of the mass step location.~When we fit for a mass-step using our analysis pipelines on simulated data, we see a bias in recovered cosmological parameters, at least in part due to the built-in assumption that the dust properties of galaxies change at $10^{10} M_{\odot}$ in the dust model for SN-Unite \citep{popovic2023dust2dust}.~We similarly do not consider alternative forms of the mass-step, such as fitting a line as a function of the host-galaxy stellar mass, or not assuming a mass-step at all, and leave such investigations to future work.

\subsection{Updating Data Release Masses to SN-Unite Masses}
\label{sec:results_cosmo_DR_masses}

In this section, we show the impact of changing our host-galaxy stellar mass measurements from the \Des~and \Pan~data-release (DR) versions, to the newly derived SN-Unite masses in this work, for (i) the full SN-Unite sample, (ii) the Unite-\Des~subsample, and (iii) the Unite-\Pan~subsample. 

\begin{table*}[htbp]
    \centering
    \caption{Flat-$w$CDM constraints for SN-Unite, Unite-DES-SN5YR, and Unite-Pantheon+ subsamples with data-release and nominal masses, without weak lensing corrections.~Reported values correspond to the medians of the marginalized posterior distributions, with 68.27\% credible intervals. For each fit we report the $\chi^2$ at the maximum likelihood, $\chi^2_{\rm ML}$ and the number of degrees of freedom. $\Delta \Omega_m$, $\Delta w $, and $\Delta Q_H$ are computed with respect to the \textbf{SN-Unite Nominal masses} constraints.~We also show constraints from 3 of the currently most widely used SN Ia samples: DES-Dovekie, Union3.1 and \Pan.\footnote{For DES-Dovekie and \Pan, we compute the medians of the marginalized posterior distributions using their data-release chains. \Pan~constraints shown here are therefore slightly different from Table 3 of \citet{brout2022pantheon+}, which uses the \textit{means} of the marginalized posterior distributions. Since Union3.1 cosmology chains are not yet public, we use values from Table 3 of \citet{rubin2026union3.1} without uncertainties for $Q_H$.} Note that apart from DES-Dovekie, which essentially shares the same dataset and methodology as the Unite-\Des~subsample in this work, the external datasets differ in both the methodology and sample compared to the Unite-\Pan~subsample.}
    \label{tab:fwcdm_mass_comparison}
    \begin{tabular}{lccccccc}
        \cline{1-8}
        \textbf{Sample (no lensing corrections)} & $\Omega_\mathrm{m}$ & $\Delta\Omega_\mathrm{m}$ & $w$ & $\Delta w$ & $Q_H$ & $\Delta Q_H$ & $\chi_{\rm ML}^2/N_{\rm dof}$ \\
        \cline{1-8}
        \multicolumn{8}{l}{\textbf{SN-Unite}} \\
        \cline{1-8}
        DR masses & $0.259^{+0.050}_{-0.060}$ & +0.053 & $-0.888^{+0.112}_{-0.119}$ & -0.110 & $-0.429^{+0.022}_{-0.022}$ & -0.011 & 2711 / 2851 \\
        \textbf{Nominal masses} & $0.206^{+0.056}_{-0.058}$ & --- & $-0.778^{+0.084}_{-0.103}$ & --- & $-0.419^{+0.021}_{-0.021}$ & --- & 2747 / 2881 \\
        \cline{1-8}
        \multicolumn{8}{l}{\textbf{DES-SN5YR subsample}} \\
        \cline{1-8}
        DR masses & $0.223^{+0.070}_{-0.071}$ & +0.017 & $-0.796^{+0.110}_{-0.142}$ & -0.018 & $-0.406^{+0.031}_{-0.030}$ & +0.013 & 1698 / 1752 \\
        Nominal masses & $0.228^{+0.069}_{-0.073}$ & +0.022 & $-0.799^{+0.112}_{-0.146}$ & -0.021 & $-0.403^{+0.031}_{-0.029}$ & +0.016 & 1698 / 1756 \\
        \cline{1-8}
        \multicolumn{8}{l}{\textbf{Pantheon+ subsample}} \\
        \cline{1-8}
        DR masses & $0.349^{+0.064}_{-0.084}$ & +0.143 & $-0.999^{+0.179}_{-0.195}$ & -0.221 & $-0.345^{+0.039}_{-0.037}$ & +0.074 & 1380 / 1483 \\
        Nominal masses & $0.243^{+0.083}_{-0.084}$ & +0.037 & $-0.778^{+0.113}_{-0.154}$ & -0.000 & $-0.356^{+0.035}_{-0.035}$ & +0.063 & 1378 / 1505 \\
        \cline{1-8}
        \multicolumn{8}{l}{\textbf{External datasets}} \\
        \cline{1-8}
        DES-Dovekie & $0.263^{+0.064}_{-0.079}$ & +0.057 & $-0.838^{+0.130}_{-0.142}$ & -0.060 & $-0.378^{+0.028}_{-0.029}$ & +0.041 & 1639 / 1817 \\
        Union3.1 & $0.240^{+0.082}_{-0.109}$ & +0.034 & $-0.776^{+0.158}_{-0.177}$ & +0.002 & $-0.363\phantom{^{+0.034}_{-0.033}}$ & +0.056 & -- / -- \\
        Pantheon+ & $0.294^{+0.061}_{-0.074}$ & +0.088 & $-0.906^{+0.140}_{-0.155}$ & -0.129 & $-0.380^{+0.034}_{-0.033}$ & +0.038 & -- / -- \\
        \cline{1-8}
    \end{tabular}
\end{table*}

\begin{figure*}
    \centering
    \includegraphics[width=0.33\linewidth]{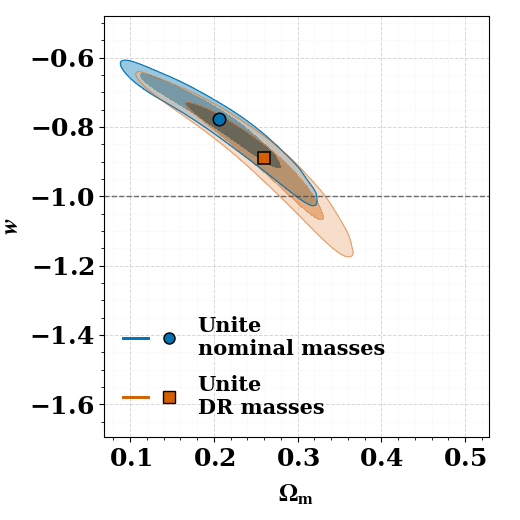}\includegraphics[width=0.33\linewidth]{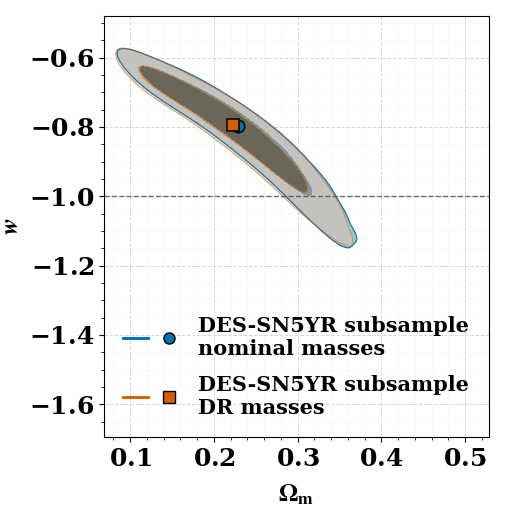}\includegraphics[width=0.33\linewidth]{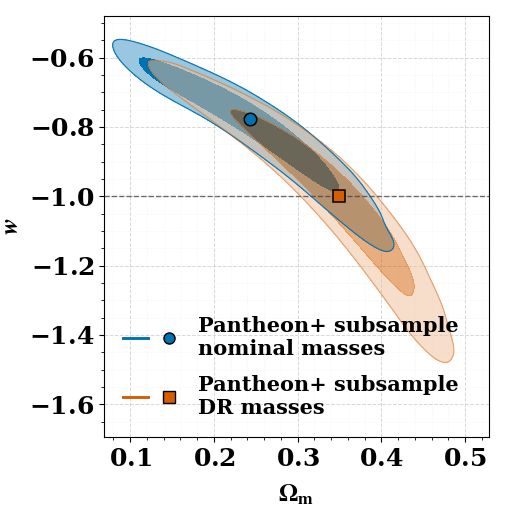}
    \caption{1$\sigma$ and 2$\sigma$ $\Omega_m $ - $w$ contours in Flat$w$CDM for DR masses updated to SN-Unite masses for the 3 samples: SN-Unite, Unite-\Des~subsample and the Unite-\Pan~subsample. The $\circ$ and $\square$ symbols denote the medians of the marginalized posterior distributions. As shown in Table~\ref{tab:fwcdm_mass_comparison}, updating the DR masses to the newly derived SN-Unite masses does not have an impact on the Unite-\Des~subsample, but shifts the Unite-\Pan~subsample and SN-Unite sample away from Flat$\Lambda$CDM, mostly along the degeneracy direction.}
    \label{fig:Unite_vs_DR_masses_FwCDM_SN_Only}
\end{figure*}

\subsubsection{Flat$w$CDM SN Ia Only}
\label{sec:results_cosmo_DR_wCDM}

In Table~\ref{tab:fwcdm_mass_comparison} and Figure~\ref{fig:Unite_vs_DR_masses_FwCDM_SN_Only}, we show the shifts in cosmological parameters for Flat$w$CDM when we update DR masses to the SN-Unite masses presented in this work for the 3 samples, SN-Unite, Unite-\Des~subsample and the Unite-\Pan~subsample as well as for the 3 most widely used external SN Ia samples: DES-Dovekie \citep{dovekie2025}, Union3.1 \citep{rubin2026union3.1}, and \Pan~\citep{brout2022pantheon+}.~The shifts $\Delta \Omega_m$, $\Delta w$, and $\Delta Q_H$ are computed with respect to the SN-Unite Nominal masses constraints.

We first see that the Unite-\Des~subsample constraints change minimally when we update the DR masses to the SN-Unite masses, which we expect given the consistency of the \Des~mass measurements with SN-Unite in Figure~\ref{fig:Unite_vs_DR_masses}.~In contrast, the Unite-\Pan~subsample constraints shift considerably in the $\Omega_m$-$w$ space, from $(\Omega_m, w) = (0.349^{+0.064}_{-0.084}, -0.999^{+0.179}_{-0.195}) $ to $(\Omega_m, w) = (0.243^{+0.083}_{-0.084}, -0.778^{+0.113}_{-0.154})$, when we update the DR masses to the SN-Unite masses.~This shift is away from a cosmological constant, with $Q_H$ shifting by about 0.3$\sigma$ due to updating the masses.~Updating the Unite-\Pan~subsample masses results in an improved consistency with the SN-Unite and Unite-\Des~subsample constraints, not only in terms of $\Omega_m$ and $w$, but also in $Q_H$.~Updating the DR masses to the new SN-Unite masses changes $Q_H$ for the full SN-Unite sample by about 0.5$\sigma$ and $w$ shifts away from a cosmological constant, as with the \Pan~subsample. 

We comment on the consistency between SN-Unite, the Unite-\Des~subsample and the Unite-\Pan~subsample, as well as compared to the external datasets shown in Table~\ref{tab:fwcdm_mass_comparison}.~First, the SN-Unite, Unite-\Des~subsample and the Unite-\Pan~subsample using nominal masses are all consistent to within 1$\sigma$ in terms of $\Omega_m$ and $w$. The Unite-\Pan~subsample is the least consistent with the SN-Unite constraints (for $\Omega_m$ and hence $Q_H$), which is expected given that the Unite-\Pan~subsample in this work is slightly different from the \Pan~SNe Ia included in SN-Unite as outlined in Section~\ref{sec:methodology_dr_masses}. 

Comparing the Unite-\Des~subsample to the external DES-Dovekie results, we find good agreement in terms of $Q_H$ ($-0.403^{+0.031}_{-0.029}$ for the Unite-\Des~subsample and $-0.378^{+0.028}_{-0.029}$ for DES-Dovekie), although there are some shifts along the degeneracy direction for $\Omega_m$ and $w$.~When we compare the Unite-\Pan~subsample (Nominal masses) to Union3.1 and \Pan, we also find reasonable agreement in terms of $Q_H$ ($-0.356^{+0.035}_{-0.035}$ for the Unite-\Pan~subsample, $-0.363$ for Union3.1, and  $-0.380^{+0.034}_{-0.033}$ for \Pan), while \Pan~shows shifts along the degeneracy direction for $\Omega_m$ and $w$ compared to the Unite-\Pan~subsample.~The comparison of the $Q_H$ values (at $z = 0.2$ as mentioned in Section~\ref{sec:results_cosmo}) suggests that all of the datasets, SN-Unite, Unite-\Des, Unite-\Pan, and the external datasets are broadly consistent with each other.~Nevertheless, using consistent host-galaxy stellar mass measurements throughout all redshifts for the official \Pan~constraints likely results in a shift mostly along the degeneracy direction \textit{away} from Flat$\Lambda$CDM. One additional interesting feature we note is the highly consistent Flat$w$CDM constraints for $\Omega_m$, $w$, and $Q_H$ between the Unite-\Pan~subsample and the Union3.1 constraints; the Unite-\Pan~subsample and Union3.1 sample share a similar dataset (with Union3.1 having approximately 40\% more SNe Ia) but use different analysis pipelines.

\subsubsection{Flat$w_0w_a$CDM SN Ia + CMB + BAO}
\label{sec:results_cosmo_DR_w0waCDM}

\begin{table*}
    \centering
    \caption{Flat-$w_0w_a$CDM constraints and preference relative to Flat-$\Lambda$CDM for SN-Unite, the Unite-DES-SN5YR, and Unite-Pantheon+ subsamples combined with BAO and CMB with data-release (DR) and nominal masses, without weak lensing corrections. Reported values correspond to the medians of the marginalized posterior distributions, with 68.27\% credible intervals. For each fit we report the $\chi^2$ at the maximum likelihood, $\chi^2_{\rm ML}$, $\Delta\chi^2_{\rm ML}$, $\Delta\chi^2_{\rm MAP}$ from MAP log-likelihood values, and the number of SNe Ia. Both significances are computed using the Wilks-theorem conversion with two degrees of freedom. We also show external constraints from 3 of the currently most widely used SN Ia samples: DES-Dovekie, Union3.1 and \Pan, as in Table~\ref{tab:fwcdm_mass_comparison}.}
    \label{tab:w0wa_preference}
    \begin{tabular}{lccccccccccc}
        \cline{1-6} \cline{8-12}
        \textbf{Dataset} & $N_{\rm SN}$ & $\Omega_\mathrm{m}$ & $w_0$ & $w_a$ & $\chi^2_{\rm ML}$ & & $\Delta\chi^2_{\rm ML}$ & $n\sigma_{\rm ML}$ & $\Delta\chi^2_{\rm MAP}$ & $n\sigma_{\rm MAP}$ & $\Delta{\rm log}\,\mathcal{Z}$ \\
        \cline{1-6} \cline{8-12}
        \multicolumn{6}{l}{\textbf{Unite (no lensing correction) + BAO + CMB}} & & & & & & \\
        \cline{1-6} \cline{8-12}
        DR masses & 2854 & $0.3032^{+0.0044}_{-0.0043}$ & $-0.902^{+0.044}_{-0.043}$ & $-0.46^{+0.17}_{-0.17}$ & 3384 &  & -7.1 & 2.2 & -8.7 & 2.5 & 2.01 \\
        Nominal masses & 2884 & $0.3051^{+0.0044}_{-0.0041}$ & $-0.864^{+0.043}_{-0.043}$ & $-0.60^{+0.17}_{-0.18}$ & 3420 &  & -11.9 & 3.0 & -13.0 & 3.2 & -0.15 \\
        \cline{1-6} \cline{8-12}
        \multicolumn{6}{l}{\textbf{DES-SN5YR subsample + BAO + CMB}} & & & & & & \\
        \cline{1-6} \cline{8-12}
        DR masses & 1755 & $0.3091^{+0.0057}_{-0.0057}$ & $-0.831^{+0.059}_{-0.059}$ & $-0.66^{+0.20}_{-0.22}$ & 2369 &  & -10.8 & 2.8 & -11.9 & 3.0 & 0.02 \\
        Nominal masses & 1759 & $0.3092^{+0.0056}_{-0.0055}$ & $-0.829^{+0.058}_{-0.058}$ & $-0.67^{+0.21}_{-0.21}$ & 2368 &  & -11.2 & 2.9 & -12.3 & 3.1 & -0.19 \\
        \cline{1-6} \cline{8-12}
        \multicolumn{6}{l}{\textbf{Pantheon+ subsample + BAO + CMB}} & & & & & & \\
        \cline{1-6} \cline{8-12}
        DR masses & 1486 & $0.3130^{+0.0056}_{-0.0055}$ & $-0.819^{+0.053}_{-0.051}$ & $-0.63^{+0.18}_{-0.19}$ & 2052 &  & -13.1 & 3.2 & -14.3 & 3.4 & -1.07 \\
        Nominal masses & 1508 & $0.3146^{+0.0055}_{-0.0053}$ & $-0.788^{+0.051}_{-0.050}$ & $-0.74^{+0.19}_{-0.19}$ & 2048 &  & -18.2 & 3.9 & -19.3 & 4.0 & -3.62 \\
        \cline{1-6} \cline{8-12}
        \multicolumn{12}{l}{\textbf{External datasets + BAO + CMB}} \\
        \cline{1-6} \cline{8-12}
        DES-Dovekie & 1820 & $0.313^{+0.005}_{-0.005}$ & $ -0.803^{+0.054}_{-0.054}$ & $ -0.72^{+0.21}_{-0.21} $ & 2231 &  & -13.5 & 3.2 & -14.2 & 3.3 & -1.7 \\
        Union3.1 & 2085 & -- & $-0.735^{+0.081}_{-0.081}$ & $-0.92^{+0.29}_{-0.25}$ & -- &  & -- & -- & -14.0 & 3.3 & -- \\
        Pantheon+ & 1701 & $0.3114^{+0.0057}_{-0.0057}$ & $ -0.838^{+0.055}_{-0.055}$ & $ -0.62^{+0.22}_{-0.19} $ & -- &  & -- & -- & -10.7 & 2.8 & -- \\
        \cline{1-6} \cline{8-12}
    \end{tabular}
\end{table*}

\begin{figure*}
    \centering
    \includegraphics[width=0.33\linewidth]{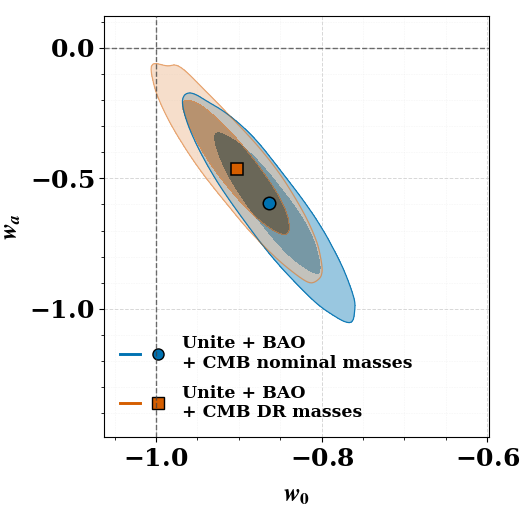}\includegraphics[width=0.33\linewidth]{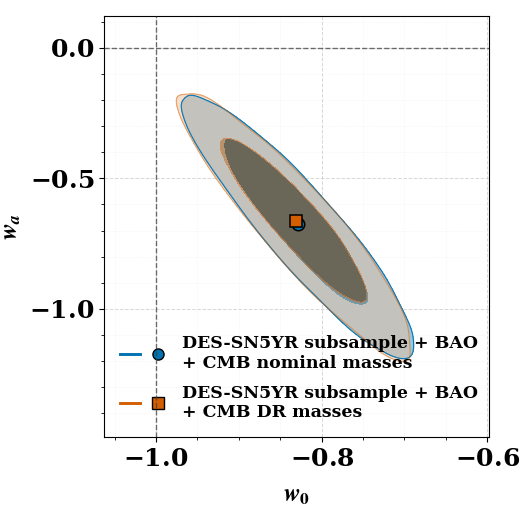}
    \includegraphics[width=0.33\linewidth]{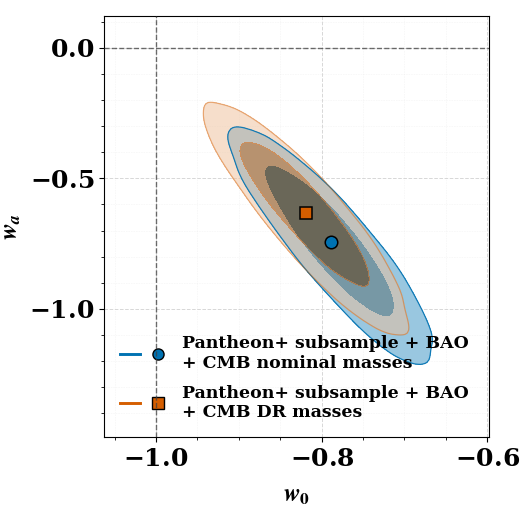}
    \caption{1$\sigma$ and 2$\sigma$ $w_0 $ - $w_a$ contours in Flat$w_0w_a$CDM for DR masses updated to SN-Unite masses for the 3 samples combined with BAO and CMB. The $\circ$ and $\square$ symbols denote the medians of the marginalized posterior distributions. As shown in Table~\ref{tab:w0wa_preference}, updating the DR masses to the newly derived SN-Unite masses does not have an impact on the \Des~subsample, but shifts the \Pan~subsample and SN-Unite samples away from Flat$\Lambda$CDM.}
    \label{fig:Flatw0waCDM_contours_DR}
\end{figure*}

In Table~\ref{tab:w0wa_preference} and Figure~\ref{fig:Flatw0waCDM_contours_DR}, we show the shifts in cosmological constraints in Flat$w_0w_a$CDM combined with BAO and CMB when we update data-release masses to the SN-Unite masses presented in this work for the 3 samples as well as for the 3 external datasets as in Section~\ref{sec:results_cosmo_DR_wCDM}.~BAO and CMB datasets are the same as in \citet{SN-Unite} and the preference for Flat$w_0w_a$CDM over Flat$\Lambda$CDM is computed in the same way as in \citet{SN-Unite} using the $\Delta \chi^2$ at the maximum likelihood (ML), the \textit{maximum a posteriori} (MAP), and $\Delta \log \mathcal Z$.~For the external datasets Union3.1 and \Pan, we use Flat$w_0w_a$CDM constraints given in \citet{hoyt2026union3.1_hosts} (Table 4, UNITY1.8) and \citet{DESI_DR2_Cosmology}.~The BAO and CMB datasets combined in SN-Unite are slightly different from those in DESI DR2; see Section 2.4 of \citet{SN-Unite} for details. 

As with Flat$w$CDM constraints presented in Section~\ref{sec:results_cosmo_DR_wCDM}, updating the DR masses to SN-Unite masses has virtually no impact on the Unite-\Des~subsample + BAO + CMB, as expected given the consistency between the \Des~and SN-Unite mass measurements. Updating the Unite-\Pan~subsample + BAO + CMB with SN-Unite masses, however, results in a shift \textit{away} from Flat$\Lambda$CDM, from 3.4$\sigma$ (3.2 $\sigma$) to 4.0$\sigma$ (3.9$\sigma$) using $\Delta \chi_{\rm MAP}^2$ ($\Delta \chi_{\rm ML}^2$). Consequently, updating the Unite + BAO + CMB sample with SN-Unite masses results in a shift \textit{away} from Flat$\Lambda$CDM, from 2.5$\sigma$ (2.2 $\sigma$) to 3.2$\sigma$ (3.0$\sigma$) using $\Delta \chi_{\rm MAP}^2$ ($\Delta \chi_{\rm ML}^2$). 

Similar to Section~\ref{sec:results_cosmo_DR_wCDM} for Flat$w$CDM, we comment on the consistency between the Unite and Unite subsamples + BAO + CMB, and with external datasets + BAO + CMB. Interestingly, we see slight shifts \textit{towards} Flat$\Lambda$CDM in the degeneracy direction as we go from Unite-\Pan~subsample + BAO + CMB $\rightarrow$ Unite-\Des~subsample + BAO + CMB $\rightarrow$ SN-Unite + BAO + CMB (Nominal masses) in Figure~\ref{fig:Flatw0waCDM_contours_DR}. 

In terms of preference for Flat$w_0w_a$CDM over Flat$\Lambda$CDM however, we see a similar significance for Unite-\Des~subsample + BAO + CMB and Unite + BAO + CMB, at 3.1$\sigma$ and 3.2$\sigma$ using $\Delta \chi_{\rm MAP}^2$ (2.9$\sigma$ and 3.0$\sigma$  using $\Delta \chi_{\rm ML}^2$, respectively) and similar (weakly preferred) $\Delta \log \mathcal Z$. 

The Unite-\Pan~subsample + BAO + CMB constraints (Nominal masses) are slightly different from the nominal Unite + BAO + CMB constraints in terms of $\Omega_m$, $w_0$, and $w_a$, and display the highest deviation from Flat$\Lambda$CDM, at 4.0$\sigma$ (3.9$\sigma$) using $\Delta \chi_{\rm MAP}^2$ ($\Delta \chi_{\rm ML}^2$) and $\Delta \log \mathcal{Z} = -3.62$.~While this may be surprising, we note that the Unite-\Pan~subsample is slightly different from the SNe Ia directly overlapping with SN-Unite due to the treatment of the spectroscopic DES-SN3YR sample. 

Comparing the Unite-\Des~subsample + BAO + CMB constraints with the DES-Dovekie + BAO + CMB constraints, we see good agreement across all cosmological parameters and preference for Flat$w_0w_a$CDM over Flat$\Lambda$CDM.

The Unite-\Pan~subsample + BAO + CMB constraints are consistent with Union3.1 + BAO + CMB as well, which is encouraging given that almost all SNe Ia in the Unite-\Pan~subsample are included in Union3.1, although the uncertainties are larger for Union3.1 + BAO + CMB, likely due to the different analysis framework. The Unite-\Pan~subsample (DR masses) + BAO + CMB  constraints are also consistent with the \Pan~+ BAO + CMB constraints.~The slight differences likely arise from analysis updates made when incorporating the \Pan~sample into SN-Unite.

This suggests that updating the \Pan~DR masses to SN-Unite masses results in an $\sim 0.6\sigma$ increase of preference for Flat$w_0 w_a$CDM over Flat$\Lambda$CDM for \Pan~+BAO+CMB, consistent with the findings of \citet{hoyt2026union3.1_hosts}. 

\subsection{Importance of Consistent Host-Galaxy Stellar Mass Measurements}
\label{sec:results_importance_mass}

The analysis in Sections~\ref{sec:results_cosmo_ana_choice} and \ref{sec:results_cosmo_DR_masses} can be summarized as follows.~While the specific host-galaxy stellar mass measurement choices have minimal impact on the SN-Unite cosmological constraints, it is crucial to maintain a consistent measurement pipeline across all of the SN Ia sample considered for cosmology.~This was true for the \Des/DES-Dovekie and Union3.1 samples, while combining various different host-galaxy mass measurements with different underlying assumptions has introduced systematic offsets in both the official \Pan~mass measurements and their cosmological constraints.

\section{Discussion and Conclusion} \label{sec:concl}

In this work, we remeasure over 98\% of the host-galaxy stellar masses in the SN-Unite Hubble Diagram using consistent photometry and SED fitting techniques. We assess the impact of photometry and SED fitting analysis variants on the mass measurements as well as cosmological constraints.~We find that our mass measurements and cosmological constraints are robust to slight changes in the mass measurement methodology, with shifts in Flat$w$CDM being at most $0.1\sigma$ in terms of $w$ and $0.3\sigma$ in terms of $Q_H$, a parameter designed to assess the cosmological shift perpendicular to the degeneracy direction in the $\Omega_m$-$w$ plane. 

Additionally, we find that our new SN-Unite masses are consistent with \Des~data-release (DR) masses, but discrepant from the \Pan~DR masses with the SN-Unite masses being higher at $z_{\rm HD} < 0.15$ and lower than \Pan~DR masses at $z_{\rm HD} \ge 0.15$ due to an internal discrepancy in the \Pan~measurements.~We computed the shifts in cosmological constraints due to updating the \Pan~and \Des~DR masses to the newly derived SN-Unite masses in Flat$w$CDM for SN Ia only, and in Flat$w_0 w_a$CDM for SN Ia + BAO + CMB. 

The Unite-\Des~subsample is not affected by the updated masses, while the Unite-\Pan~subsample constraints move \textit{away} from Flat$\Lambda$CDM when the new SN-Unite masses are used, confirming the findings of \citet{vincenzi2025response,hoyt2026union3.1_hosts}.~The preference for Flat$w_0w_a$CDM over Flat$\Lambda$CDM when the Unite-\Pan~subsample is combined with BAO and the CMB increases from 3.4$\sigma$ to 4.0$\sigma$ using the $\Delta \chi_{\rm MAP}^2$ (3.2$\sigma$ to 3.9$\sigma$ using the $\Delta \chi_{\rm ML}^2$), highlighting the importance of consistent host-galaxy stellar mass measurements across the full SN Ia sample considered.

This analysis is not the end of investigations of host-galaxy correlations with SN Ia brightnesses post light-curve corrections.~As briefly outlined in Section~\ref{sec:intro}, future work should strive to find a more physically motivated, yet reliable method to account for host-galaxy correlations with SN Ia luminosities, as opposed to using the global host-galaxy stellar mass. One promising pathway may be the two-stretch population approach shown in \citet{rubin2026union3.1}.~In addition, future work should fit for the host-galaxy mass-step location, rather than fixing it to $10^{10}M_{\odot}$, as done in this work. 

Nonetheless, the analysis provided in \citet{SN-Unite}, as well as this current work, shows that the 3 representative SN Ia cosmological constraints (Union3.1, DES-Dovekie, and \Pan) agree with each other, and combining each of them with BAO and CMB points to a $\sim 3\sigma$ preference of time-evolving dark energy over Flat$\Lambda$CDM under a frequentist interpretation.~By re-measuring the host-galaxy stellar masses consistently throughout all of SN-Unite, we mitigate potential systematic shifts in the cosmological constraints associated with heterogeneous host-galaxy stellar mass measurements, while providing the tightest cosmological constraints to date.


\section{Data Availability}

The SN-Unite host-galaxy stellar mass measurements and photometry \textit{will} be publicly available upon acceptance of this manuscript. 


\begin{acknowledgments}

We thank Miranda Gorsuch, Michael Rosenthal, and Christy Tremonti for helpful suggestions.~JL and KB are supported by the US Department of Energy grant DE‐SC0022950.~Ryan C. is grateful for the support provided by the Big Questions Institute and its commitment to fostering fundamental research.~Parts of this research were conducted by the Australian Research Council Centre of Excellence for Gravitational Wave Discovery (project number CE230100016) funded by the Australian Government. LG acknowledges financial support from CSIC, MCIN and AEI 10.13039/501100011033 under projects PID2023-151307NB-I00, PIE 20215AT016, CEX2020-001058-M, and by the MaX-CSIC Excellence Award MaX4-SOMMA-ICE. MS was partially supported by DOE grant DE-FOA-0003177.~TEMB is funded by Horizon Europe ERC grant no. 101125877.~AM is supported by Australian Research Council award (DE230100055). We acknowledge the use of OpenAI's \verb|ChatGPT| and \verb|Codex| for efficient code implementation and manuscript proofreading.~We thank Sesh Nadathur for a question that prompted us to revisit and clarify our MAP values. In the first version submitted to arXiv, we inadvertently used the MAP posterior, which includes a contribution from the prior volume, rather than the likelihood evaluated at the MAP.

\end{acknowledgments}

\appendix

\section{Details on Host Galaxy Stellar Mass Uncertainties}
\label{sec:appendixA}

\begin{figure*}
    \centering
    \includegraphics[width=0.47\linewidth]{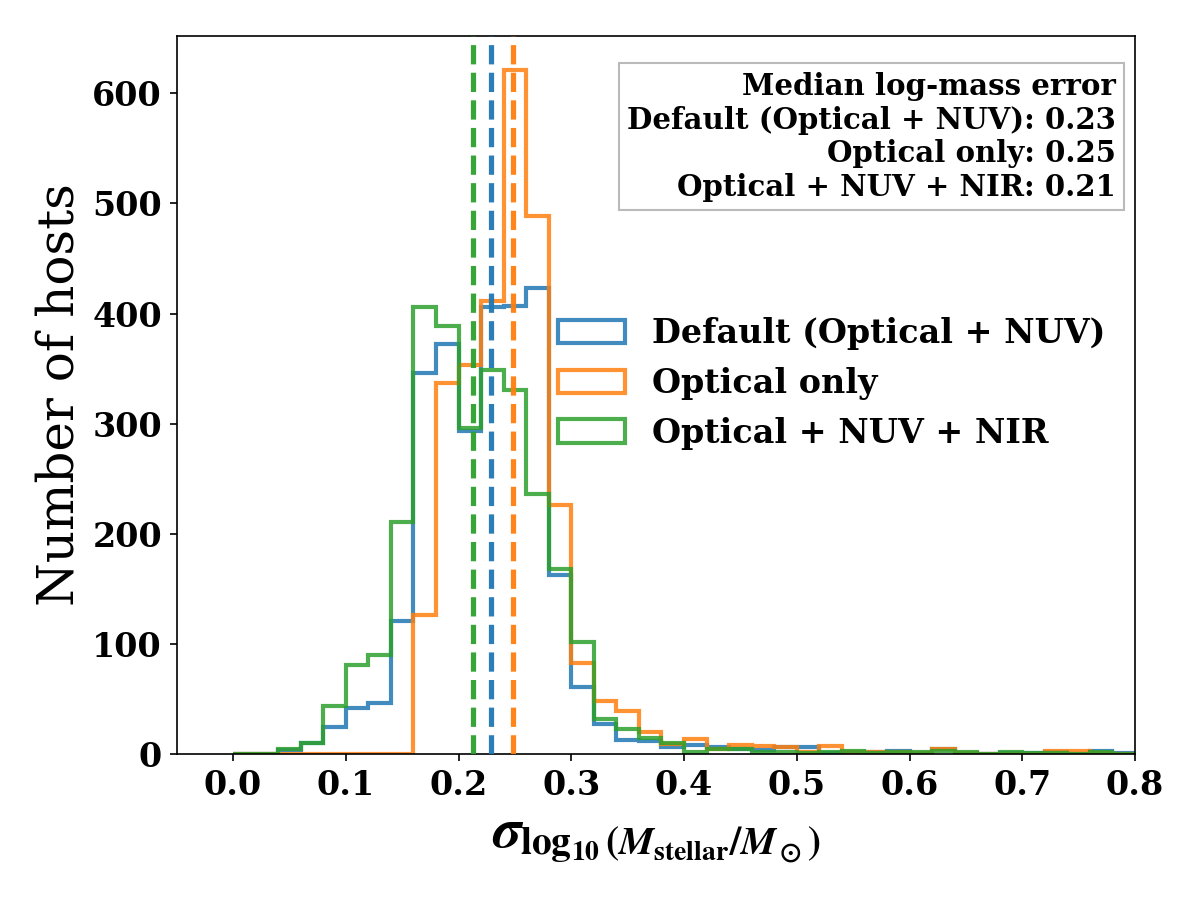}\includegraphics[width=0.47\linewidth]{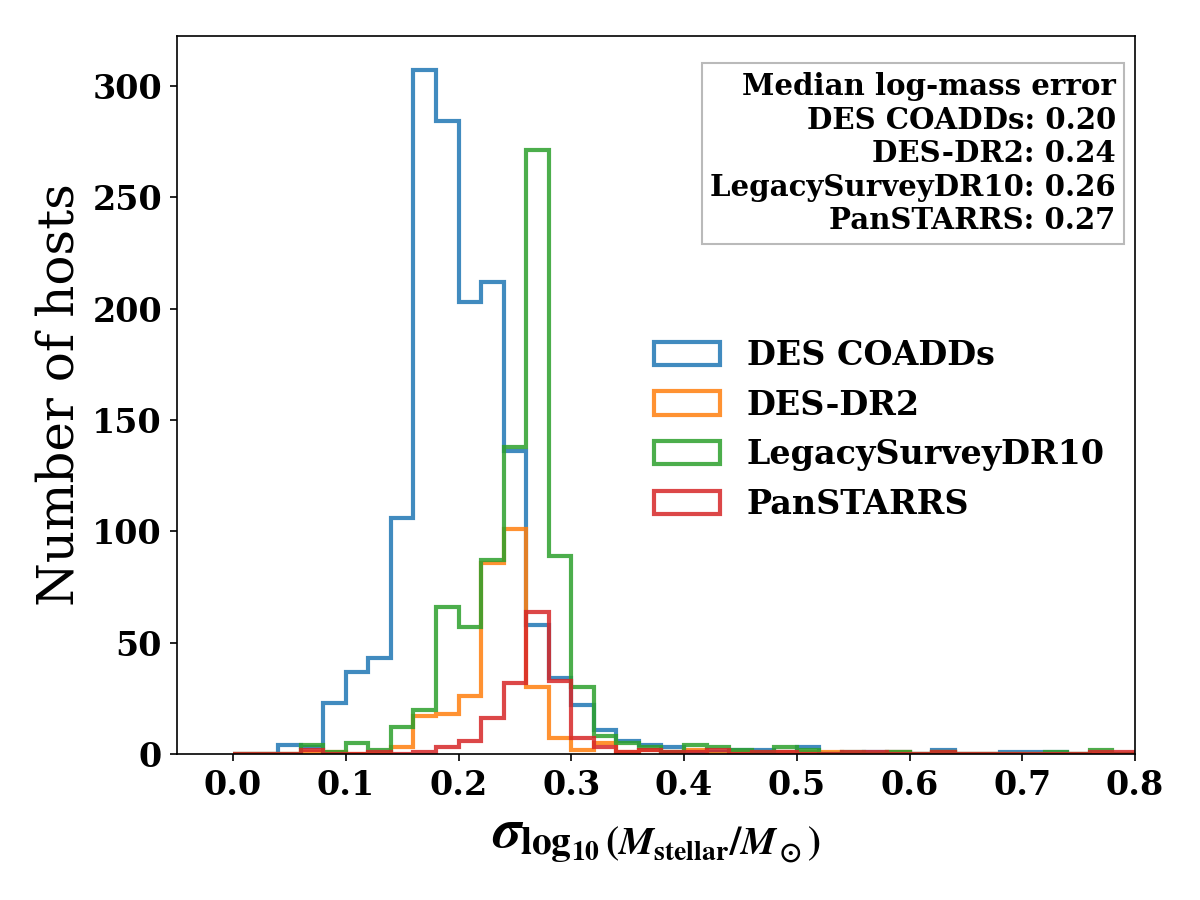}
    \caption{Histograms of the SN-Unite host-galaxy stellar mass uncertainties: (Left) based on wavelength-coverage and (Right) based on the optical survey used for photometry.~Including \textit{GALEX} NUV and \textit{2MASS} NIR photometry decreases the median host-galaxy stellar mass uncertainties.~However, we do not include \textit{2MASS} photometry as it degrades the SED fit, likely due to the low resolution of \textit{2MASS} images. The right figure shows the host-galaxy stellar mass uncertainties decreasing with deeper depth optical surveys; this is due to using the default value of 0.1 for the CIGALE `additionalerror,' which is the relative error added in quadrature for fluxes and the extensive properties.}
    \label{fig:log_mass_err_distributions}
\end{figure*}

In Figure~\ref{fig:log_mass_err_distributions}, we show the distributions of the SN-Unite host-galaxy log stellar mass uncertainties, depending on the wavelength coverage (left), and the optical survey used for photometry.~While the median uncertainty is 0.25 dex when using optical photometry only, this decreases to 0.23 dex and 0.21 dex when the NUV from \textit{GALEX} and NIR from \textit{2MASS} are included respectively. As we found that \textit{2MASS} \textit{JHKs} photometry typically degrades the SED fit rather than improves the fit due to its large ($\sim 0.8 $ mag) scatter, we only use the optical and \textit{GALEX} NUV photometry for our default host-galaxy stellar mass measurements. The right panel of  Figure~\ref{fig:log_mass_err_distributions} shows the host-galaxy stellar mass uncertainties decreasing as we go to deeper depth optical surveys, which is due to using the default value of 0.1 for the \verb|CIGALE| \verb|additionalerror|, or the relative error added in quadrature for fluxes and the extensive properties. 

We note that the default median host-galaxy log stellar mass uncertainties for SN-Unite, at 0.23 dex, is much larger than the \Des~data-release median host-galaxy log stellar mass uncertainties, which is around 0.03 dex. This is because \Des~only included statistical uncertainties arising from the uncertainties on photometry itself, while we take a more conservative approach to account for systematic (modeling) uncertainties by setting the \verb|CIGALE| \verb|additionalerror| parameter to 0.1.~We acknowledge that this is a simplistic approach, and future work can find alternative ways to quantify systematic uncertainties in host-galaxy stellar mass measurements more robustly. 

\section{Initial Mass Functions}
\label{sec:appendixB}

In this section, we provide the equations for the Initial Mass Functions (IMFs) considered in this work. In Equation~\ref{eq:IMFs}, $\xi(\log_{10} M) = \ln(10) M\xi(M)$, and we set $\xi_0 = 0.0443$ to match the Chabrier IMF normalization.

\begin{align}
    \xi_{\rm Salpeter}(\log_{10} M) 
    &= \ln(10)\,\xi_0 M^{-1.35}, \\[0.5em]
    \xi_{\rm Kroupa}(\log_{10} M) &=
    \ln(10)
    \begin{cases}
        A_1 M^{0.7}, & M < 0.08\,M_\odot, \\
        A_2 M^{-0.3}, & 0.08\,M_\odot \le M < 0.5\,M_\odot, \\
        A_3 M^{-1.3}, & M \ge 0.5\,M_\odot,
    \end{cases} \\[0.5em]
    A_3 &= \xi_0, \\
    A_2 &= A_3 \, (0.5)^{-(2.3-1.3)}, \\
    A_1 &= A_2 \, (0.08)^{-(1.3-0.3)}, \\[0.5em]
    \xi_{\rm Chabrier}(\log_{10} M) &=
    \ln(10)
    \begin{cases}
        0.158
        \exp\left[-\dfrac{\left(\log_{10} M-\log_{10} 0.079\right)^2}
        {2(0.69)^2}
        \right], & M < 1\,M_\odot, \\[1.0em]
        4.43\times10^{-2} M^{-1.3}, & M \ge 1\,M_\odot .
    \end{cases}
    \label{eq:IMFs}
\end{align}



\label{sec:pubcharge}


\bibliography{refs}{}
\bibliographystyle{aasjournalv7.1}





\end{document}

%% file: host_galaxy_author_list.tex
\author[0000-0001-6633-9793]{J.~Lee}
\affiliation{Department of Physics, University of Wisconsin Madison, WI 53706-1390, USA}\email[show]{lee2547@wisc.edu}
\author[0000-0002-7436-3950]{R.~Camilleri}
\affiliation{School of Mathematics and Physics, University of Queensland, Brisbane, QLD 4072, Australia}\email{}
\author[0000-0002-4213-8783]{T.~M.~Davis}
\affiliation{School of Mathematics and Physics, University of Queensland, Brisbane, QLD 4072, Australia}\email{}
\author[0000-0001-5402-4647]{D.~Rubin}
\affiliation{Department of Physics and Astronomy, University of Hawai‘i at Mānoa, Honolulu, Hawai‘i 96822}\email{}
\author[0000-0001-8156-0429]{K.~Bechtol}
\affiliation{Department of Physics, University of Wisconsin Madison, WI 53706-1390, USA}\email{}
\author[0000-0002-1296-6887]{L.~Galbany}
\affiliation{Institute of Space Sciences (ICE-CSIC), Campus UAB, Carrer de Can Magrans, s/n, E-08193 Barcelona, Spain}\email{}
\affiliation{Institut d'Estudis Espacials de Catalunya (IEEC), 08860 Castelldefels (Barcelona), Spain}\email{}
\author[0000-0003-2764-7093]{M.~Sako}
\affiliation{Department of Physics and Astronomy, University of Pennsylvania, Philadelphia, PA 19104, USA}\email{}
\author[0000-0001-9053-4820]{M.~Sullivan}
\affiliation{School of Physics and Astronomy, University of Southampton, Southampton, SO17 1BJ, UK}\email{}
\author[0000-0003-3939-7167]{T.~E.~Müller-Bravo}
\affiliation{School of Physics, Trinity College Dublin, The University of Dublin, Dublin 2, Ireland}
\affiliation{Instituto de Ciencias Exactas y Naturales (ICEN), Universidad Arturo Prat, Chile}\email{}
\author[0000-0002-4934-5849]{D.~Scolnic}
\affiliation{Department of Physics, Duke University, Durham, NC 27708, USA}\email{}
\author[0000-0001-8788-1688]{M.~Vincenzi}
\affiliation{Department of Physics, University of Oxford, Denys Wilkinson Building, Keble Road, Oxford OX1 3RH, United Kingdom}\email{}
\author[0000-0001-5201-8374]{D.~Brout}
\affiliation{Departments of Astronomy and Physics, Boston University, Boston, MA 02215}\email{}
\author[0000-0003-1731-0497]{C.~Lidman}
\affiliation{Research School of Astronomy and Astrophysics \& Centre for Gravitational Astrophysics, The Australian National University, Canberra, ACT, Australia}\email{}
\author[0000-0001-8211-8608]{A.~Möller}
\affiliation{Centre for Astrophysics and Supercomputing, Swinburne University of Technology, John St, Hawthorn, VIC 3122, Australia}\email{}
\author[0000-0002-8000-6642]{P.~Shah}
\affiliation{Department of Physics and Astronomy, University College London, Gower Street, London, UK}\email{}
\author[0000-0002-5389-7961]{M.~Acevedo}
\affiliation{Department of Physics, Duke University, Durham, NC 27708, USA}\email{}
\author[0000-0003-1997-3649]{P.~Armstrong}
\affiliation{The Research School of Astronomy and Astrophysics, The Australian National University, Canberra, ACT 2611, Australia}\email{}
\affiliation{Department of Physics, University of California Berkeley, Berkeley, CA 94720, USA}
\affiliation{E.O. Lawrence Berkeley National Laboratory, 1 Cyclotron Rd., Berkeley, CA 94720, USA}\email{}
\author[0000-0001-7700-1069]{B.~A.~Bassett}
\affiliation{Wits MIND Institute and School of Computer Science and Applied Mathematics, University of the Witwatersrand, Johannesburg, South Africa}
\affiliation{School for Data Science and Computational Thinking, Stellenbosch University, South Africa}\email{}
\author[0000-0003-3917-0966]{R.~C.~Chen}
\affiliation{Kavli Institute for Particle Astrophysics \& Cosmology, P. O. Box 2450, Stanford University, Stanford, CA 94035, USA}
\affiliation{SLAC National Accelerator Laboratory, Menlo Park, CA 94025, USA}
\affiliation{Department of Physics, Stanford University, 382 Via Pueblo Mall, Stanford, CA 94305, USA}\email{}
\author[0000-0002-8357-7467]{H.~T.~Diehl}
\affiliation{Fermi National Accelerator Laboratory, P. O. Box 500, Batavia, IL 60510, USA}
\affiliation{Kavli Institute for Cosmological Physics, University of Chicago, Chicago, IL 60637, USA}
\affiliation{Department of Astronomy and Astrophysics, University of Chicago, Chicago, IL 60637, USA}\email{}
\author[0000-0003-4079-3263]{J.~Frieman}
\affiliation{Fermi National Accelerator Laboratory, P. O. Box 500, Batavia, IL 60510, USA}
\affiliation{Kavli Institute for Cosmological Physics, University of Chicago, Chicago, IL 60637, USA}
\affiliation{Department of Astronomy and Astrophysics, University of Chicago, Chicago, IL 60637, USA}\email{}
\author[0009-0002-2654-843X]{K.~Grech}
\affiliation{School of Mathematics and Physics, University of Queensland, Brisbane, QLD 4072, Australia}\email{}
\author[0000-0002-8012-6978]{B.~Popovic}
\affiliation{School of Physics and Astronomy, University of Southampton, Southampton, SO17 1BJ, UK}\email{}
\author[0000-0002-8687-0669]{B.~O.~Sánchez}
\affiliation{Aix Marseille Univ, CNRS/IN2P3, CPPM, Marseille, France}\email{}
\author[0000-0002-4283-5159]{B.~E.~Tucker}
\affiliation{Research School of Astronomy and Astrophysics \& Centre for Gravitational Astrophysics, The Australian National University, Canberra, ACT, Australia}\email{}